\documentclass[aip, jcp, twocolumn, floatfix, superscriptaddress, reprint, 10pt]{revtex4-2}
\usepackage[utf8]{inputenc}
\usepackage[paperwidth=210mm,paperheight=297mm,centering,hmargin=2cm,vmargin=2.5cm]{geometry}

\usepackage[english]{babel}

\usepackage{hyperref}
\usepackage{tikz}
\usepackage{amsmath}
\usepackage{cleveref}
\usepackage{graphicx}
\usepackage{booktabs}
\usepackage{listings}
\usepackage{xspace}
\usepackage{multirow}
\usepackage[inline]{enumitem}
\usepackage{lipsum}
\usepackage{float}
\usepackage{titlesec}
\usepackage{capt-of}

\titlespacing*{\section}{0pt}{\baselineskip}{0.8\baselineskip}
\titlespacing*{\subsection}{0pt}{\baselineskip}{0.6\baselineskip}

\usepackage[normalem]{ulem}
\definecolor{tangerine}{rgb}{0.944,0.522,0}
\definecolor{verde}{rgb}{0.,0.6,0}
\definecolor{rosso}{rgb}{0.9,0.0,0.2}
\definecolor{orange}{rgb}{1.0,0.5,0.0}

\newcommand{\resub}[1]{#1}

\newcommand{\metatensor}{\texttt{metatensor}\xspace}
\newcommand{\metatomic}{\texttt{metatomic}\xspace}
\newcommand{\metatrain}{\texttt{metatrain}\xspace}
\newcommand{\cxx}{C\nolinebreak\hspace{-.05em}\raisebox{.4ex}{\tiny\bf +}\nolinebreak\hspace{-.10em}\raisebox{.4ex}{\tiny\bf +}\xspace}

\begin{document}

\title{
    \metatensor and \metatomic: foundational libraries for interoperable atomistic machine learning
}

\author{Filippo Bigi}
\thanks{These two authors contributed equally}
\author{Joseph W. Abbott}
\thanks{These two authors contributed equally}
\author{Philip Loche}
\author{Arslan Mazitov}
\author{Davide Tisi}
\author{Marcel F. Langer}
\author{Alexander Goscinski}
\author{Paolo Pegolo}
\author{Sanggyu Chong}
\author{Rohit Goswami}
\author{Pol Febrer}
\author{Sofiia Chorna}
\author{Matthias Kellner}
\author{Michele Ceriotti}
\author{Guillaume Fraux}
\email[]{guillaume.fraux@epfl.ch}

\affiliation{Laboratory of Computational Science and Modeling, Institute of Materials, École Polytechnique Fédérale de Lausanne, 1015 Lausanne, Switzerland}

\begin{abstract}
Incorporation of machine learning (ML) techniques into atomic-scale modeling has proven to be an extremely effective strategy to improve the accuracy and reduce the computational cost of simulations.
It also entails conceptual and practical challenges, as it involves combining very different mathematical foundations, as well as software ecosystems that are very well developed in their \resub{own right}, but do not share many commonalities.
To address these issues and facilitate the adoption of ML in atomistic simulations, we introduce two dedicated software libraries.
The first one, \metatensor, provides multi-platform and multi-language storage and manipulation of arrays with many potentially sparse indices, designed from the ground up for atomistic ML applications. By combining the actual values with metadata that describes their nature and that facilitates the handling of geometric information and gradients with respect to the atomic positions, \metatensor provides a common framework to enable data sharing between ML software --- typically written in Python --- and established atomistic modeling tools --- typically written in Fortran, C or \cxx.
The second library, \metatomic, provides an interface to store an atomistic ML model and metadata about this model in a portable way, facilitating the implementation, training and distribution of models, and their use across different simulation packages.
We showcase a growing ecosystem of tools, \resub{including} low-level libraries, training utilities, \resub{and} interfaces with existing software packages that demonstrate the effectiveness of \metatensor and \metatomic in bridging the gap between traditional simulation software and modern ML frameworks.
\end{abstract}

\maketitle
\setcitestyle{super}

\section{Motivation}

Machine learning (ML) techniques have profoundly transformed atomistic simulations. Their use has become so ubiquitous that it is now rare to find research papers \resub{in this domain} that do not incorporate ML, either as the primary modeling engine or in combination with physics-based approaches. This widespread adoption has spurred the development of a broad ecosystem of software libraries and tools, enabling researchers to apply ML methods to a wide range of problems in materials science and chemistry\cite{Behler2007,Bartok2010,Rupp2012,Smith2017,Butler2018,fourGenBeheler,Unke2019,Deringer2021,WANG2018178,zeng2023deepmdkit,schutt2022schnetpack,Batzner_NatCommun_2022_v13_p2453,MACEpaper}.

At the same time, the universe of ML software has become increasingly diverse, spanning a wide range of programming languages and frameworks. This includes Fortran, C, \cxx, Python (with libraries such as scikit-learn\cite{scikit-learn}, PyTorch\cite{paszke2019pytorch}, and JAX\cite{jax2018github}), and Julia\cite{bezanson2017julia}. Driven by the fast pace of innovation in ML, many of these libraries evolve rapidly, posing challenges for both developers and practitioners.
For developers, the diversity often leads to the pragmatic choice of building on a single framework -- such as NumPy\cite{numpypaper} or PyTorch -- thereby limiting the resulting software to specific languages or environments. This fragmentation reduces the reusability and composability of methods across different projects.
For practitioners, the situation is equally challenging. Keeping pace with the rapid development of new ML tools and frameworks can be overwhelming, even for experts in the field. This makes it difficult to build, deploy, contribute and extend atomistic ML software in a reliable and sustainable way.

One prominent example of this challenge is the integration of atomistic ML models with traditional simulation engines. These engines include molecular dynamics packages such as LAMMPS\cite{thompson2022lammps}, GROMACS\cite{ABRAHAM201519}, ASE\cite{larsen2017atomic}, OpenMM\cite{openmm}, and i-PI\cite{litm+24jcp}, as well as quantum chemistry and \emph{ab initio} codes like PySCF, Quantum ESPRESSO\cite{giannozzi_advanced_2017}, CP2K\cite{cp2kpaper}, and FHIaims\cite{AIMS,FHIaims2025}. ML models are also increasingly used to define collective variables in enhanced sampling tools like PLUMED\cite{TRIBELLO2014604}, and in data analysis and visualization workflows.

However, the diversity of programming languages and data structures used across engines poses a significant barrier to interoperability. Bridging ML models with simulation software often requires custom interfaces, which are time-consuming to develop and difficult to maintain. As a result, users are typically constrained to a small subset of compatible combinations -- limiting both flexibility and reproducibility in practice. These combinations are often determined by the specific interests and priorities of individual developers, rather than by the full range of scientific use cases. Consequently, practitioners may find that the available integrations do not support the particular workflow or simulation protocol required for their research, with little prospect that such support will be added unless they implement it themselves.

\begin{figure}[htbp]
\includegraphics[width=\linewidth]{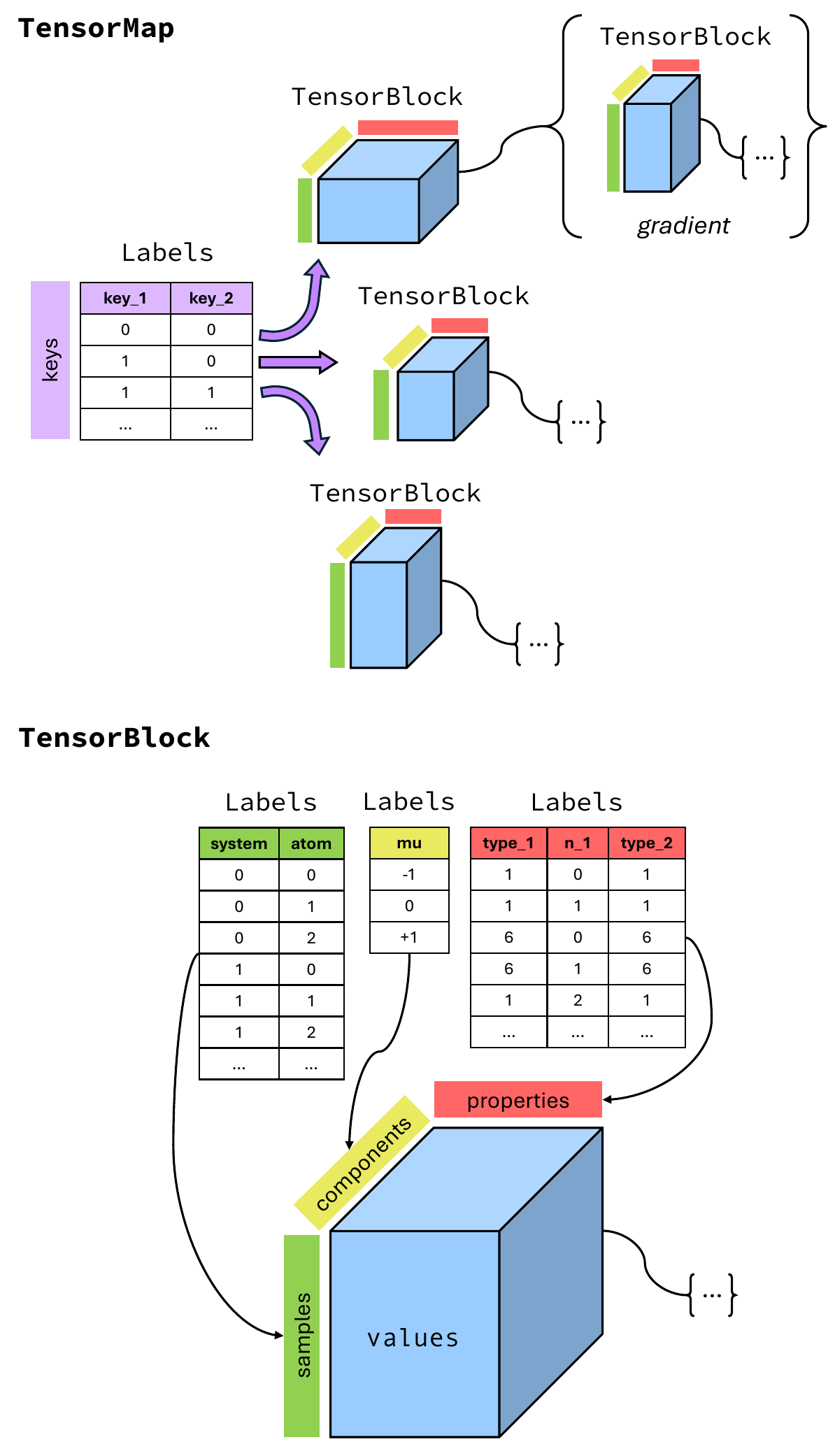}
\caption{\label{fig:mts-types}
\resub{Schematic illustration of the three primary objects in \metatensor. (1)~A \texttt{TensorMap} is a key/value map that acts as a block-sparse storage format, and forms the highest-level object. It groups multiple \texttt{TensorBlock}s and (optionally) their associated gradient \texttt{TensorBlock}s (in curly brackets) into a complete representation, each indexed by an entry in the \emph{keys} \texttt{Labels}. (2)~Each \texttt{TensorBlock} is comprised of \texttt{values} -- i.e. dense floating-point data -- and its corresponding metadata in the \emph{samples}, \emph{components}, and \emph{properties} \texttt{Labels}. (3)~\texttt{Labels} are used to store metadata, with named dimensions and unique row entries. Shown are one set of \texttt{Labels} corresponding to the keys of the \texttt{TensorMap} (with dimensions ``key\_1'' and ``key\_2''), and three set of \texttt{Labels} corresponding to the \emph{samples} (``system'' and ``atom'' dimensions), components (a single ``mu'' dimension), and \emph{properties} (``type\_1'', ``n\_1'', and ``type\_2'' dimensions) of a \texttt{TensorBlock}.}}
\end{figure}

Here, we present the \metatensor and \metatomic libraries, along with their growing ecosystem, which aim to standardize and improve the interoperability and usability of machine learning software with traditional tools for atomistic modeling and simulation. Our libraries are compatible with multiple platforms, programming languages, and ecosystems to support a broad range of applications.

\metatensor facilitates the exchange of \emph{data} between software packages by annotating arrays with metadata that makes them self-describing. It also supports storing gradients and their associated metadata alongside the data itself -- an essential feature for many atomic-scale quantities of interest, such as forces and polarizabilities. In addition, \metatensor’s metadata structure enables compact representations of advanced sparsity patterns, which are common in atomistic ML and significantly reduce memory usage.

\metatomic, in turn, enables the exchange of \emph{code} between software packages. This is critical because ML models consist not only of numerical data (the “weights”) but also of code -- often serialized, compiled, or framework-specific -- that defines how inputs are transformed into outputs. By standardizing the representation of atomistic models, \metatomic makes it easy to share models and run them across a wide range of simulation tools. This offers a major advantage: As outlined above, each ML architecture must implement multiple interfaces for different simulators, often in different languages. A shared format for inputs and outputs allows model developers to focus on architecture design, rather than the repetitive task of writing and maintaining interfaces.

The \metatensor and \metatomic ecosystem is already widely adopted, with many research projects using them\cite{lopa+23prm, mazi+24jpm,loch+25jcp,cign+24acscs,chon+25fd, turk2025reconstructions,mazitov2025petmad,suman2025,bigi2025flashmd}, and new libraries across a wide range of modeling tasks, from high-performance numerical kernels to user-friendly model training interfaces\cite{litm+24jcp,TRIBELLO2014604,Lin2025}.

In the following, we present the core design of \metatensor and \metatomic, focusing on their data formats and implementation details, along with basic usage examples. We then describe the broader ecosystem built around them and highlight the external tools that now support \resub{these} libraries. We conclude with a perspective on the evolving atomistic ML software landscape and future directions for both libraries and their ecosystems.

\section{Sharing data with metatensor}

\subsection{Data format}

\metatensor introduces a self-describing sparse array data format which is generally applicable but incorporates many features that are typically relevant for atomistic ML. This format encapsulates both the arrays themselves and their associated metadata, facilitating efficient storage and manipulation of complex scientific data. At its core, \metatensor introduces three primary objects (see Figure~\ref{fig:mts-types}).

(1) \texttt{Labels} define the metadata attached to different data entries. It is a set of named, multi-dimensional indices, which one can visualize as a 2-dimensional array of integers, each column representing one named dimension, and each row representing one data entry.

(2) A \texttt{TensorBlock} contains a single multi-dimensional data array, together with its associated metadata and gradients. Each dimension of the array is decorated with metadata stored in \texttt{Labels}, the first dimension, called \emph{samples}, describing ``which object this data pertains to'' and the last dimension, called \emph{properties}, describing ``what properties of the object have been stored''. For example, when using \metatensor to store the electron density on a set of atomic basis functions \resub{for a single system,}~the samples would describe the atom index, and the properties the different basis functions onto which the electron density has been projected. Further intermediate dimensions of the array are used to handle tensorial data of arbitrary rank, and are called \emph{components}. For example, when storing forces, the three components of the force vector (along the x, y, and z axes) are stored as \emph{components} of the \texttt{TensorBlock}.

\texttt{TensorBlock} can also contain gradients of the data, in the form of (potentially multiple) additional  \texttt{TensorBlock} instances, each labeled by a string describing the gradient parameter. As an example, when storing the energy of a system, one can have a gradient with respect to \texttt{"positions"}, which would be the negative of the forces, and a gradient with respect to the cell \texttt{"strain"}, which would contain the negative of the virial. As gradients are simply \texttt{TensorBlock} instances, with their own data, increasingly higher-order gradients of gradients can be easily stored in a recursive way. This close integration of data stored with their gradients simplifies access patterns and reduces the risk of data/gradient mismatching.

(3) Finally, a \texttt{TensorMap} can be used to store related data present in multiple \texttt{TensorBlock} objects. This need can arise in two ways. First, when data is sparse in a well-defined manner, one can group together data that has the same patterns in a \texttt{TensorBlock} and use multiple blocks to store the whole data, avoiding the need to store data which is identically zero. Similarly, sometimes different parts of the information need a different set of tensorial \emph{components} (e.g. for spherical tensors) or different set of \emph{properties} (e.g. when representing the electron density each atom type can use a different set of basis functions). A \texttt{TensorMap} can be seen as a key/value map where all the values are \texttt{TensorBlock}, and the keys \resub{are tuples of integers corresponding to entries in a \texttt{Labels} object. In other words, a \texttt{TensorMap} implements a block sparse storage format, with each block stored as a \texttt{TensorBlock}, each with (optionally) associated gradient \texttt{TensorBlock}s. The use of \texttt{Labels} for both the block keys and indices for the dense axes of each block provides transparent annotation of potentially complicated data formats. Specific examples are explored further in Section~\ref{sec:usage_examples}}.

For data exchange and long-term storage, \metatensor includes robust serialization and deserialization functionality \resub{using the} \texttt{npz} format from NumPy to preserve both array data and metadata. The \texttt{npz} file format is a \texttt{zip} file containing multiple arrays in \texttt{npy} format. This format is language-agnostic and easy to implement, ensuring that the data will still be readable even if \metatensor no longer exists. \resub{This ease of implementation was the main reason for us to pick the \texttt{npz} format above other serialization formats such as NetCDF\cite{Rew1990} or HDF5\cite{HDF5}, but this does not prevent future version of the code to offer alternative serialization of the in-memory \metatensor data.}

\subsection{Usage examples}\label{sec:usage_examples}

\begin{figure*}[htbp]
    \includegraphics[width=\linewidth]{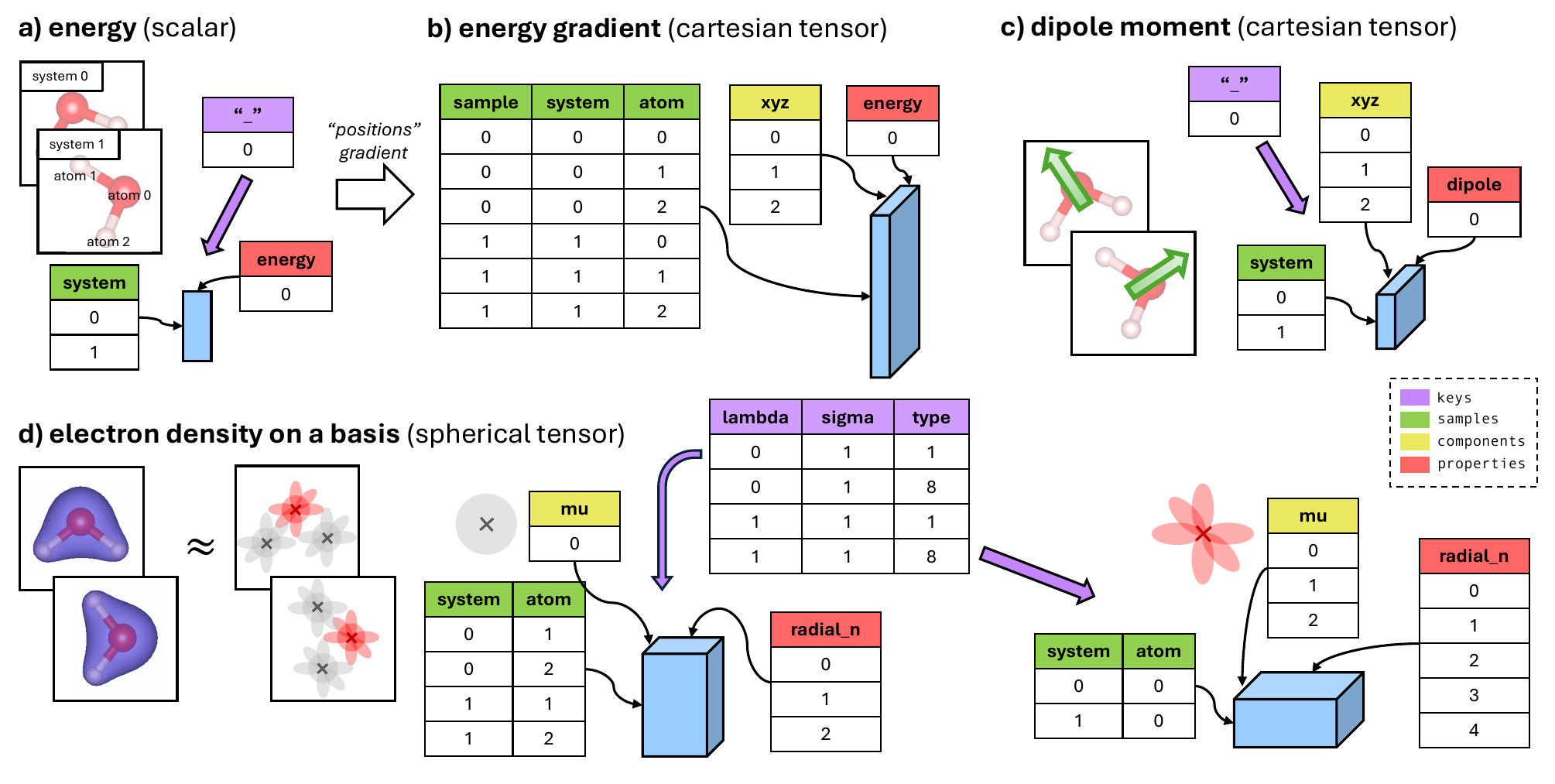}
    \caption{Examples of data stored in \texttt{TensorMap} objects. a) The atomization energy, a scalar, is stored in a single \texttt{TensorBlock}. b) The gradient of the energy with respect to the atomic positions, a per-atom Cartesian tensor, can be stored in a gradient \texttt{TensorBlock} associated with the energy block from a). c) The molecular dipole moment, a per-system Cartesian tensor, is also stored in a single block. d) The atom-centered basis decomposition of the electron density, a spherical tensor on an atomic basis, can be stored using the \texttt{TensorMap} block-sparse format with each block corresponding to irreducible representations of the O(3) group for each atom type.}
    \label{fig:mts-tensor-examples}
\end{figure*}

While \metatensor imposes some structure on how the data is stored and annotated, it is flexible enough to cover many use cases involving atomistic data. We'll explore some of these  atomistic use cases here, and show how they can be mapped to \metatensor data storage objects.

\textbf{Scalar properties} are the simplest kind of data, \resub{being invariant under rotations}. For example, the energy of a system (Figure~\ref{fig:mts-tensor-examples}a) can be stored in a \texttt{TensorBlock} with the following metadata. The \textit{samples} contain a single dimension that tracks the ``system'' index, with batches of data for different systems being stored in tensors stacked along this dimension. No \textit{components} are required for a scalar, and the \textit{properties} contain a single ``energy'' dimension with a single index value, typically \texttt{0}. \resub{As there is no specific block sparsity pattern for the energy that would benefit from a multi-block representation, the single \texttt{TensorBlock} is stored in a \texttt{TensorMap} indexed by a placeholder key \texttt{"\_"}.}

\textbf{Cartesian tensors} typically have \textit{components} axes that store Cartesian directions, or products of them, depending on the rank of the tensor. One example is the energy gradient with respect to atomic positions (Figure~\ref{fig:mts-tensor-examples}b), corresponding to the negative of the atomic forces. In the \textit{samples}, the ``system'' and ``atom'' dimensions index the atoms in each system, while the ``sample'' dimension tracks the sample in the original block (i.e., the energy in Figure~\ref{fig:mts-tensor-examples}a) for which we have the gradient. The \textit{components} have values \texttt{[0, 1, 2]} representing the x, y, and z components of the vector, and the \textit{properties} are the same as for the energy block. Another example is the per-system Cartesian dipole moment (see Figure \ref{fig:mts-tensor-examples}b). This can be stored in a single \texttt{TensorBlock}. The samples track the system indices, a single component axis tracks the ``xyz'' components of the vector, and the properties are labelled with a single ``dipole'' dimension.

\textbf{Spherical tensors:} For observables expressed on a spherical basis, the targets are expressed in irreducible representations or \textit{irreps} of the O(3) group that are labeled by their angular order ``lambda'' and \resub{inversion} parity ``sigma''. A simple example is the dipole moment, this time on the spherical basis, consisting of a single irrep labeled by \texttt{[lambda=1, sigma=1]}. A more complex example is a scalar field, such as the electron density, decomposed onto an atom-centered basis set\cite{gris+18prl,gris+19acscs,FHIaims2025} (Figure~\ref{fig:mts-tensor-examples}c). The target is stored in blocks that correspond to its irreps, and can also be stored as block sparse in the atom type, for which different radial basis definitions exist. Other electronic structure targets on a spherical basis, such as the Hamiltonian\cite{niga+22jcp,cign+24acscs,suman2025} or density matrix can also be stored efficiently with \texttt{metatensor}.

\subsection{Notes on the implementation}

The core \metatensor library is implemented in Rust, and it exposes a C application programming interface (API) to the external world. Most programming languages are able to call a C API, allowing users from diverse ecosystems to access the same functionality without duplicating implementation. In particular, we provide bindings for \cxx, Python, Rust, and TorchScript to the C API. TorchScript is the language used by PyTorch\cite{pasz+19nips} to export models defined in Python, making the models usable from \cxx directly, no longer requiring a Python interpreter. Binding \metatensor to TorchScript is done in the \texttt{metatensor-torch} \cxx library and Python package, allowing users to define custom models in Python and execute them inside C and \cxx simulation tools. See Section~\ref{sec:metatomic} for more information about running models.

Internally, \metatensor handles the storage for all metadata itself, but delegates the storage for actual data to downstream code. From the library perspective, all the data is stored as opaque array pointers with a handful of functions that can operate on them. \metatensor thus allows the storage and usage of any kind of data, including arrays that live in GPU memory. Users of the library are then expected to provide custom implementations of this array interface tailored to their own needs and data storage preferences. For Python users, we provide integrations for the NumPy \texttt{ndarray} and PyTorch \texttt{Tensor} classes. The latter is fully integrated with PyTorch's automatic differentiation framework, enabling seamless gradient tracking and manipulation.

\resub{While libraries for labeled data already exist (for example Xarray\cite{Hoyer2017} or Pandas\cite{Mckinney2010}), they lack explicit support for gradient storage, which is essential in physical sciences where gradients with respect to inputs represent physical quantities -- for example forces and virial as the negative gradients of the energy with respect to atomic positions and strain respectively. Most of these libraries are also hard to use from arbitrary programming languages, and difficult to integrate with ML frameworks. Thus, \metatensor was born of the need for a unified framework that can seamlessly handle gradients and integrate with other ML frameworks. Finally, the explicit block-sparse structure of \metatensor makes it a good fit for storing equivariant data --- which is especially useful in atomistic machine learning --- and is hard to replicate within existing libraries.}

In addition to the core data structures, \metatensor provides a rich set of utilities in the form of companion packages such as \texttt{metatensor-operations} and \texttt{metatensor-learn}. The former includes functions for manipulating and transforming tensor data, such as filtering, joining, or performing linear algebra operations, while the latter offers high-level abstractions for training ML models using the \texttt{metatensor} format. In the next sections, we will explore both \texttt{metatensor-operations} and \texttt{metatensor-learn} and give some examples of how they can be used to define custom models.

\subsection{metatensor-operations}

\begin{figure*}
    \includegraphics[width=\linewidth]{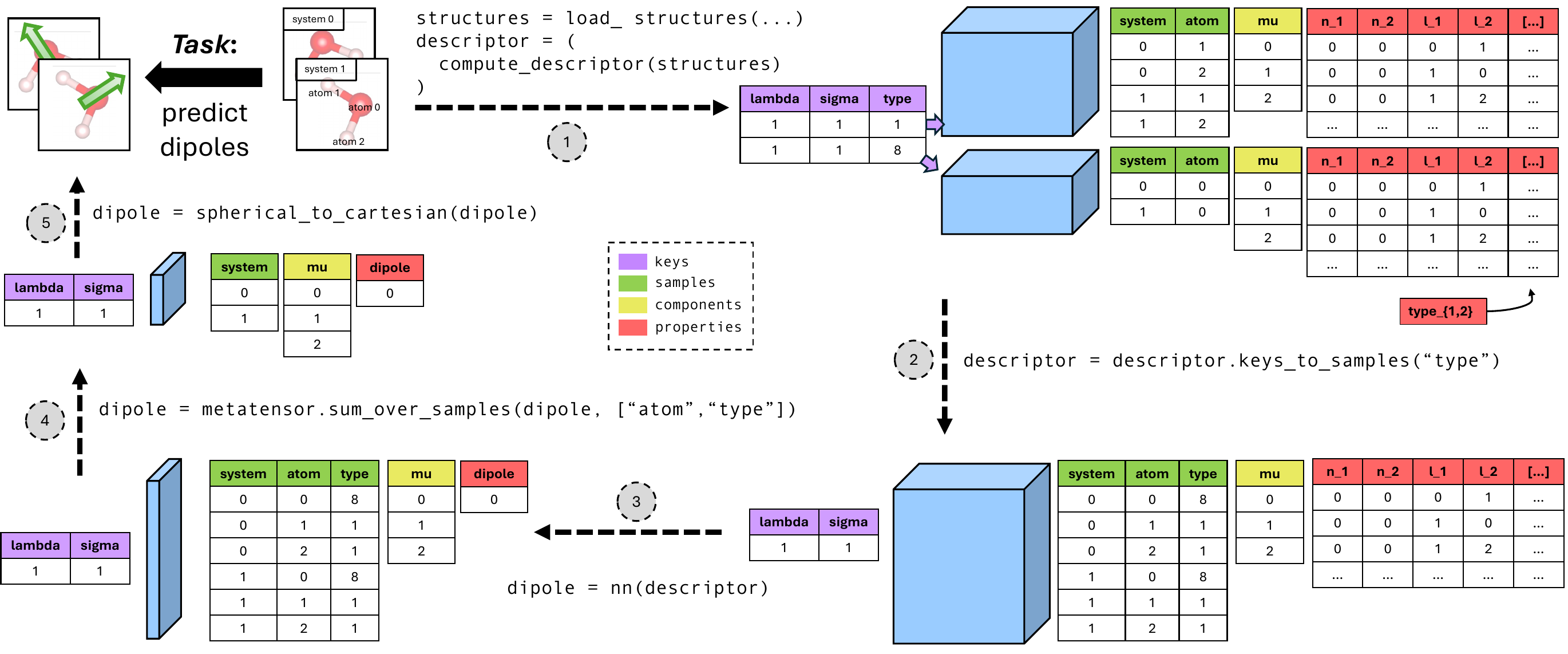}
    \caption{An example of a series of transformations using \texttt{metatensor-operations} and \texttt{metatensor-learn}, for the task of transforming an equivariant descriptor into a predicted molecular dipole moment. (1) An equivariant descriptor is computed for some input systems (or ``frames''), for example a $\lambda$-SOAP using the \texttt{featomic.EquivariantPowerSpectrum} calculator, see Section~\ref{sec:featomic}. The descriptor is separated into blocks matching the different O(3) symmetries of the target property (\texttt{lambda=1} and \texttt{sigma=1}) and the central atom \texttt{type}. (2) \texttt{keys\_to\_samples} is used to group together blocks with the same symmetries, making the descriptor dense along the atom \texttt{type}. (3) A neural network created using \texttt{metatensor-learn} utilities transforms the features of the descriptor into a per-atom dipole prediction. (4) \texttt{sum\_over\_samples} is used to aggregate local predictions to form the per-structure total dipole prediction. (5) the prediction is transformed from the spherical to the Cartesian basis.}
    \label{fig:mts-prediction-example}
\end{figure*}

\texttt{metatensor-operations} includes a number of low-level operations for processing and manipulating data and their gradients stored in \metatensor format, while also automatically keeping track of how the corresponding metadata transforms. These operations can be seamlessly chained together to construct complex data transformation pipelines, in the context of both the processing/analysis of atomistic data and the construction of custom ML models. Crucially, operations can use both automatic differentiation frameworks and explicit forward gradient propagation, making sure the gradients of relevant properties are available when needed.

These operations are provided as a Python package, and compatible with both the standard Python bindings to metatensor as well as the TorchScript bindings, allowing one to use the operations either directly from Python or through TorchScript inside simulation engines written in C, \cxx and other compiled languages. \resub{Defining the operations in a Python package instead of at the Rust level allows us to implement them using functions from NumPy and PyTorch directly, greatly reducing the work required to add new operations and immediately making the operations run on GPU when using PyTorch for data storage.}

Many operations are available, from mathematical functionalities such as \texttt{metatensor.multiply} or \texttt{metatensor.pow}, to logic functions such as \texttt{metatensor.allclose}, and creation operations such as \texttt{metatensor.zeros\_like}. A complete list of the available operations is available in the documentation at \url{https://docs.metatensor.org/latest/operations/}. Their API and behavior are similar to analogous operations found in \texttt{NumPy} or \texttt{PyTorch}. Mirroring the philosophy of \metatensor, these operations behave consistently regardless of the underlying data type used to store arrays inside \texttt{TensorMap}, whether it's a NumPy array or a PyTorch tensor. In practice, the implementation leverages the appropriate NumPy or PyTorch operations, with the code automatically dispatching to the correct backend based on the data type. This can be easily extended to incorporate new array backends such as JAX or CuPy in the future.

More complex operations are also available, performing metadata-aware transformations of the sparse data. For example, \texttt{metatensor.sum\_over\_samples} allows reduction over one or more dimensions along the samples axis. One common use case for this operation is the summation of local (or ``per-atom'') energy contributions of a model prediction to get per-system predictions. Other examples include operations to slice arrays according to a metadata filter (\texttt{metatensor.slice}), join multiple \texttt{TensorMap} while accounting for block-sparsity patterns (\texttt{metatensor.join}), or  manipulate the metadata structure (\texttt{metatensor.filter\_blocks} and \texttt{metatensor.sort}).

\subsection{metatensor-learn}

\texttt{metatensor-learn} provides high-level abstractions for building ML models and training workflows in the \texttt{metatensor-learn} Python package. Here, we give tools that mimic the API of PyTorch's \texttt{nn} module and data loading tools, but adapted to work with \metatensor data types. This ensures that users already familiar with the above APIs can use \texttt{metatensor-learn} directly, and quickly adapt existing workflows and model definitions. As in the case of \texttt{metatensor-operations}, the building blocks available in the \texttt{metatensor-learn} module are fully compatible with TorchScript and automatic differentiation, for the same reasons.

Simple neural networks can be constructed using building blocks such as \texttt{Linear}, \texttt{ReLU}, \emph{etc}. Each of these modules defines a specific transformation (such as a linear map or an activation function) transforming \texttt{TensorMap} objects with proper handling of metadata and sparsity. Furthermore, \texttt{metatensor-learn} features a more general \texttt{ModuleMap} class that can be used to define complex neural networks composed of arbitrary layers, each with custom per-block transformations.

We also provide symmetry-preserving building blocks to allow the construction of O(3)-equivariant models. These apply per-block transformations whose functional forms depend on the irreducible representation of the input \texttt{TensorMap}. For example, the transformation of an equivariant descriptor into predicted atomic dipoles shown in Figure~\ref{fig:mts-prediction-example} can be achieved with an equivariance-preserving neural network, such as \texttt{nn = Sequential[EquivariantLinear(...), InvariantReLU(...), EquivariantLinear(...)]}, with an appropriate choice of input and output dimension sizes.

Finally, \texttt{metatensor-learn} contains utilities for loading data and creating batches for gradient-descent optimization, with a similar API to PyTorch's \texttt{Dataset} and \texttt{DataLoader}. Overall, the tools in \texttt{metatensor-learn} allow one to create custom complex models and training workflows that operate on \metatensor data types. A complete list of building blocks provided is available in the documentation at \url{https://docs.metatensor.org/latest/learn/}.

\section{Sharing ML models with metatomic}
\label{sec:metatomic}

Machine learning provides powerful and efficient methods to predict the properties of atomistic systems. However, using ML models in practice -- especially within molecular simulations -- can be challenging, particularly for researchers who are not experts in ML.
A typical atomistic study requires two components: a way to compute properties (which an ML model can provide), and a way to sample the appropriate thermodynamic ensemble, such as molecular dynamics or Monte Carlo. For both tasks, a wide variety of tools and approaches exist. Over time, many different ML models and simulation engines have been developed independently. However, using them together is often difficult: the lack of standard interfaces means that only a small subset of models and engines are compatible. Creating and maintaining connections between each ML model and each simulation tool requires substantial effort, which limits practical interoperability.

\begin{figure}
\begin{lstlisting}[language=Python]
class ModelWrapper(mts.learn.nn.Module):
    def __init__(self, model):
        super().__init__()
        self.model = model
        self.nl_options = NeighborListOptions(...)

    def requested_neighbor_lists(self):
        return [self.nl]

    def forward(
        self,
        systems: List[System],
        outputs: Dict[str, ModelOutput],
        selected_atoms: Optional[Labels]
    ) -> Dict[str, TensorMap]:
        if "energy" not in outputs:
            return {}

        if selected_atoms is not None:
            raise NotImplementedError()

        energies = []
        for system in systems:
            neighbors = system.get_neighbor_list(
                self.nl_options
            ).block()

            # run the model
            energy = self.model(
                positions=system.positions,
                types=system.types,
                cell=system.cell,
                pair_idx=neighbors.samples.values,
                pair_vec=neighbors.values,
            )
            energies.append(energy)

        # put the output(s) in metatensor
        block = TensorBlock(
            values=torch.vstack(energies),
            samples=samples,
            components=[],
            properties=Labels("energy", [[0]]),
        )

        return {
            "energy": TensorMap(
                Labels("_", [[0]]), [block]
            )
        }

# Define metadata and export
wrapped = ModelWrapper(torch.load("model.pt"))
atomistic = AtomisticModel(
    wrapped,
    ModelMetadata(...),
    ModelCapabilities(...),
)
atomistic.save("metatomic_model.pt")
\end{lstlisting}
\caption{\resub{Minimal pseudo-code example showing how to wrap a pretrained  PyTorch model and export it as a \metatomic model. The wrapped model requests a neighbor list that will be provided by the simulation engine.}}
\label{fig:metatomic-export}
\end{figure}

\metatomic addresses this challenge by defining a standardized interface between ML models and simulation engines. This interface specifies a set of functions and data types for communication between the two sides. By implementing this common interface, the integration effort is reduced from $\mathcal{O}(M \times N)$ to $\mathcal{O}(M + N)$, \resub{where $M$ is the number of ML models and $N$ the number of simulation engines.} Once an ML model supports the \metatomic interface, it can be used with any simulation engine that also supports it --- and vice versa. This follows a classic “hourglass” design in software engineering, where a narrow, stable interface enables a large variety of components above and below.

The \metatomic interface can be understood in terms of three key steps. First (1), the model declares what it is capable of computing. Most importantly, it provides a list of named outputs -- such as energies, forces, dipole moments, or ML-based features -- that define its predictive capabilities. This allows a single model to support multiple tasks. Second (2), the model may request additional data from the simulation engine beyond standard inputs like atomic species, positions, and the simulation cell. In particular, it can request one or more neighbor lists, which are provided by the engine. This design enables the reuse of fast, optimized neighbor list routines already present in many simulation packages. Future versions of \metatomic will extend this mechanism to support other quantities, such as atomic charges or spins. Finally (3), the simulation engine asks the model to compute one or more of its declared outputs for a given system. The request can also be restricted to a subset of atoms, which enables use cases such as hybrid simulations (machine learning and classical force fields used together) or computing collective variables within localized regions of a system. \resub{A minimal pseudo-code example demonstrating this workflow is provided in Figure~\ref{fig:metatomic-export}.}

\begin{figure}[htbp]
\centering
\includegraphics[width=\linewidth]{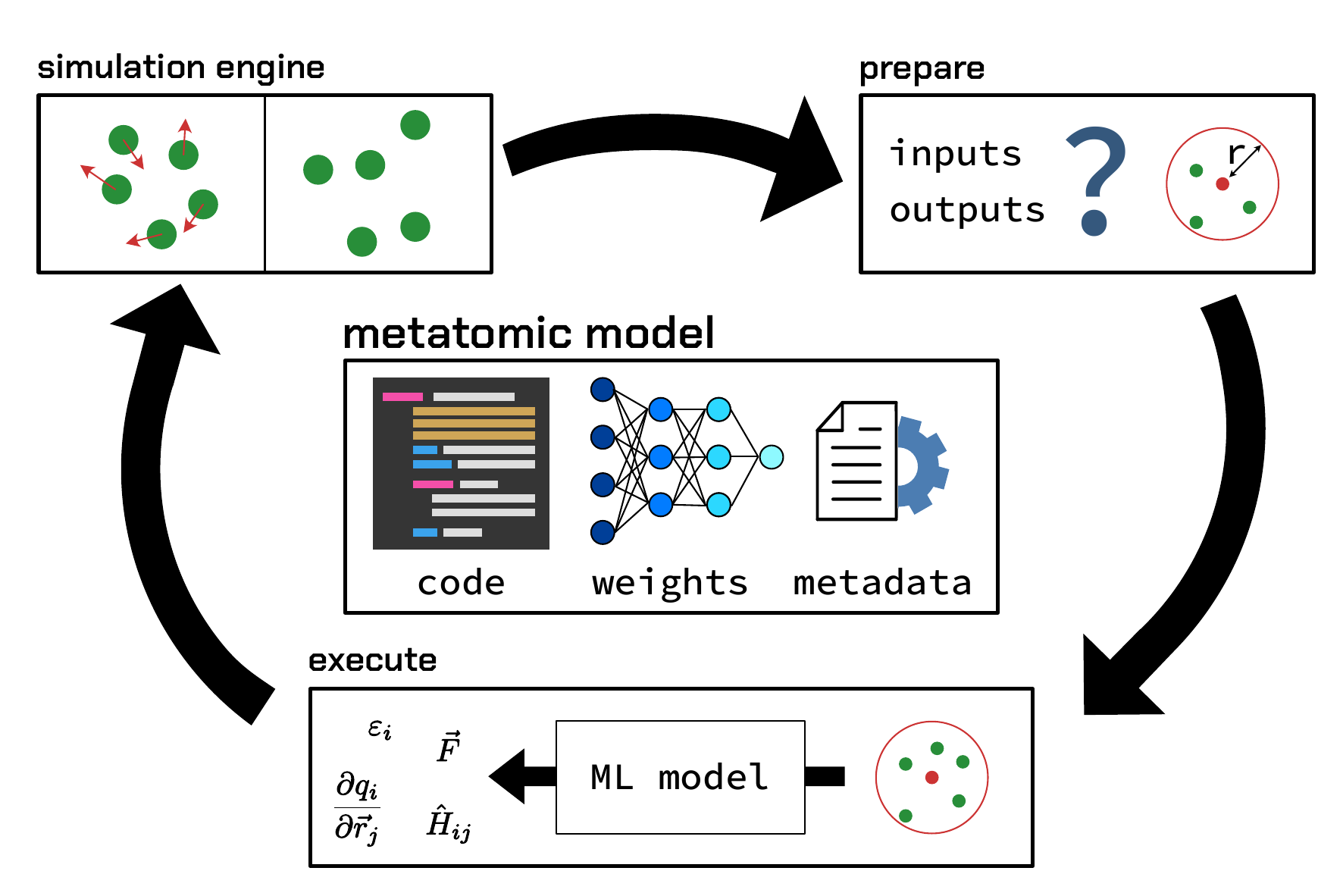}
\caption{Schematic illustration of the information flow between the ML model and simulation engine in \metatomic. The model contains the serialized code, the weights of the ML model, metadata about the model itself (authors, article references) and metadata about what the model can do. A simulation engine will query the model about what inputs it requires --- including any required neighbor lists --- and which outputs it can produce. The engine can then prepare the inputs and run the model to get the output using a unified interface.}
\label{fig:metatomic}
\end{figure}

\metatomic does not impose constraints on the content or structure of model outputs. In practice, outputs are returned as \texttt{TensorMap} objects from \metatensor, allowing them to represent any data relevant to communication between model and engine. To promote consistency, \metatomic defines a set of standardized output names -- such as ``\texttt{energy}'' or ``\texttt{features}'' -- along with associated metadata conventions. Models can also return custom outputs when needed. Overall, \metatomic is flexible enough to support a wide range of tasks in atomistic modeling, while remaining simple enough to implement on both sides of the interface.

A list of simulation engines currently supporting \metatomic is given in Section~\ref{sec:engines}. ML models conforming to the interface can be created in various ways -- for instance, using our training tool \metatrain (Section~\ref{sec:metatrain}), which enables training and fine-tuning of models without writing additional code.

\resub{To evaluate the overhead added by the \metatomic interface, we ran simulations of liquid water using the same MACE-OFF24-medium potential\cite{Kovacs2023, Batatia2022} through both the \texttt{pair\_style mace} integration with LAMMPS, and \metatomic's integration with LAMMPS. We used the KOKKOS variant of both interfaces, and ran simulations for 100 steps after a warmup of 30 steps on an H100 NVIDIA GPU. For 128 water molecules, \texttt{pair\_style mace} could execute 18.7~timestep/s while \metatomic could execute 18.3~timestep/s. For 1024 water molecules, \texttt{pair\_style mace} could execute 2.58~timestep/s while \metatomic could execute 2.53~timestep/s. Overall, this shows that the overhead introduced by \metatomic is negligible, of the order of 2~$\mu s$/atom when the model execution takes around 130~$\mu s$/atom. We picked \texttt{pair\_style mace} over other integrations such as \texttt{pair\_style mliap} or \texttt{symmetrix}\cite{Kovacs2023} because it uses the same underlying technology (TorchScript interpreter in \cxx) as \metatomic. These alternative implementations can often be faster than TorchScript, and we are planning to make sure they can also be used with \metatomic.}

\section{A modular ecosystem for atomistic ML}

Building on our foundational libraries, we introduce a suite of software packages for atomistic machine learning, each targeting different users and levels of abstraction. The resulting ecosystem is modular, enabling users to assemble components from different libraries into fully customized models. This modularity is made possible by the well-defined interfaces provided by \metatensor and \metatomic, which allow software packages to interoperate without requiring knowledge of each other’s internal implementation.

\subsection{metatrain}
\label{sec:metatrain}

\metatrain is a command-line tool for training and evaluating ML models for atomistic systems. It standardizes the training process for a wide range of state-of-the-art ML architectures and atomistic targets, offering users maximum flexibility in model development and deployment.

The interface revolves around three core commands: (1) \texttt{mtt train}, which trains a model using a user-defined \texttt{options.yaml} configuration; (2) \texttt{mtt export}, which converts trained checkpoints to \metatomic models for deployment with simulation engines; and (3) \texttt{mtt eval}, which evaluates an exported model on a user-supplied dataset.

\begin{figure}[htbp]
\includegraphics[width=\linewidth]{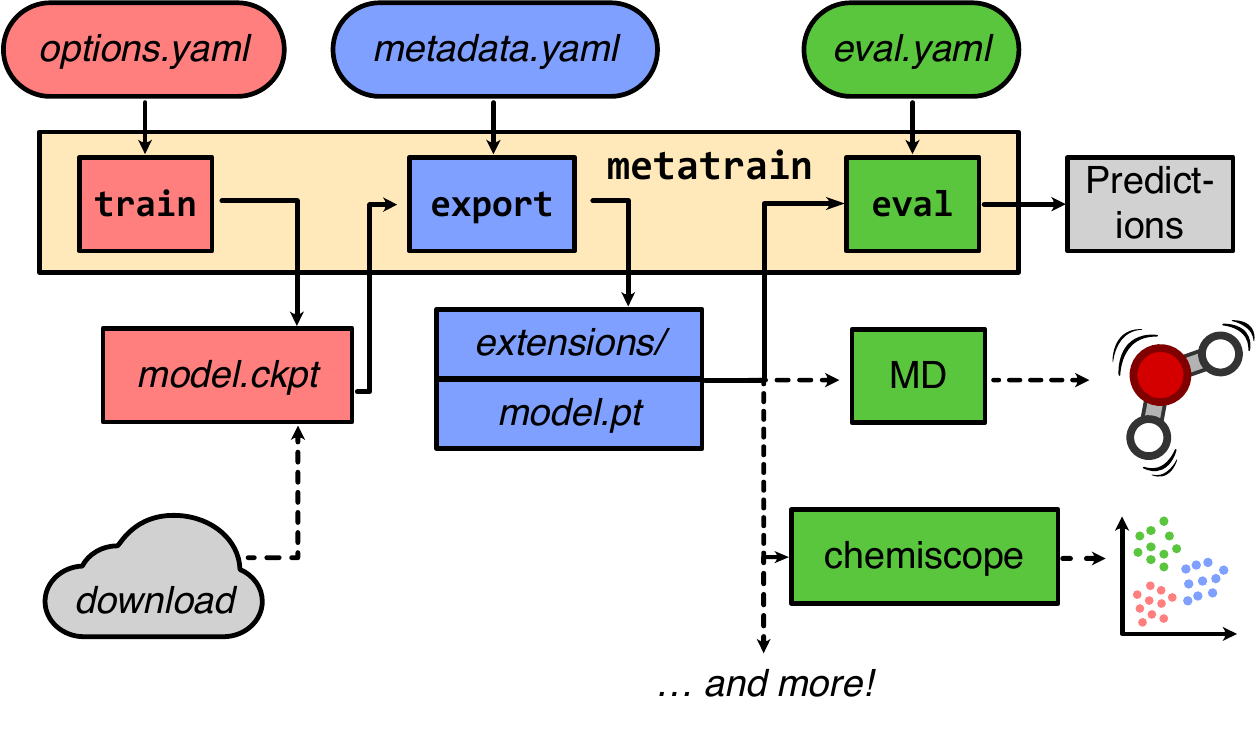}
\caption{\label{fig:mtt-design}
A schematic representation of a workflow using \metatrain commands \texttt{train}, \texttt{export}, and \texttt{eval}. The \texttt{model.ckpt} is a model-specific checkpoint, and the exported \texttt{model.pt} contains a \metatomic-compatible model. In addition to evaluation within \metatrain, exported models can be used with simulation engines, for dataset exploration in chemiscope, and more.}
\end{figure}

The training and deployment process shown in \autoref{fig:mtt-design} is highly customizable. The only strict requirement is that inputs must be representable as atomic-like data --- i.e. decorated (and possibly periodic) point clouds in three-dimensional space. This minimal constraint supports a wide range of \resub{per-system and per-atom learning targets, including interatomic potentials, the electronic density of states\cite{benm+20prb,how+25prm}, NMR chemical shieldings\cite{kellnerDeepLearningModel2025a} and electron densities\cite{gris+19acscs}. Targets defined for pairs of atoms such as the Hamiltonian\cite{niga+22jcp,cign+24acscs,suman2025} and density matrix are under active development.} All options can be configured directly via the YAML file, without requiring custom code. This allows practitioners to benchmark different architectures for a given application by simply adjusting the relevant section of the configuration file.

\metatrain supports a broad range of ML architectures, from classical models like GAP\cite{Bartok2010} and Behler-Parrinello neural networks\cite{Behler2007}, to modern graph neural networks, including both invariant and equivariant variants\cite{pozd-ceri23nips}. All models adhere to a unified interface, which allows seamless integration of additional features such as long-range corrections or uncertainty quantification. For uncertainty estimates, last-layer rigidity-based methods\cite{bigi+24mlst,chon+25fd} have been implemented, and a shallow ensemble approach\cite{kell-ceri24mlst} will be added in future releases. Thanks to the standardized interface, these extensions are immediately compatible with all architectures in \metatrain.

\metatrain also scales efficiently across multiple GPUs and supports on-the-fly loading of large datasets from disk. Several models have already been trained and published using \metatrain, including the PET-MAD general-purpose interatomic potential\cite{mazitov2025petmad}, the chemical shift predictor ShiftML\cite{kellnerDeepLearningModel2025a}, and the FlashMD models for accelerated molecular dynamics\cite{bigi2025flashmd}. Externally developed models, such as MACE~\cite{MACEpaper}, are also supported.

\subsection{featomic}
\label{sec:featomic}

\begin{figure*}
    \centering
    \includegraphics[width=\linewidth]{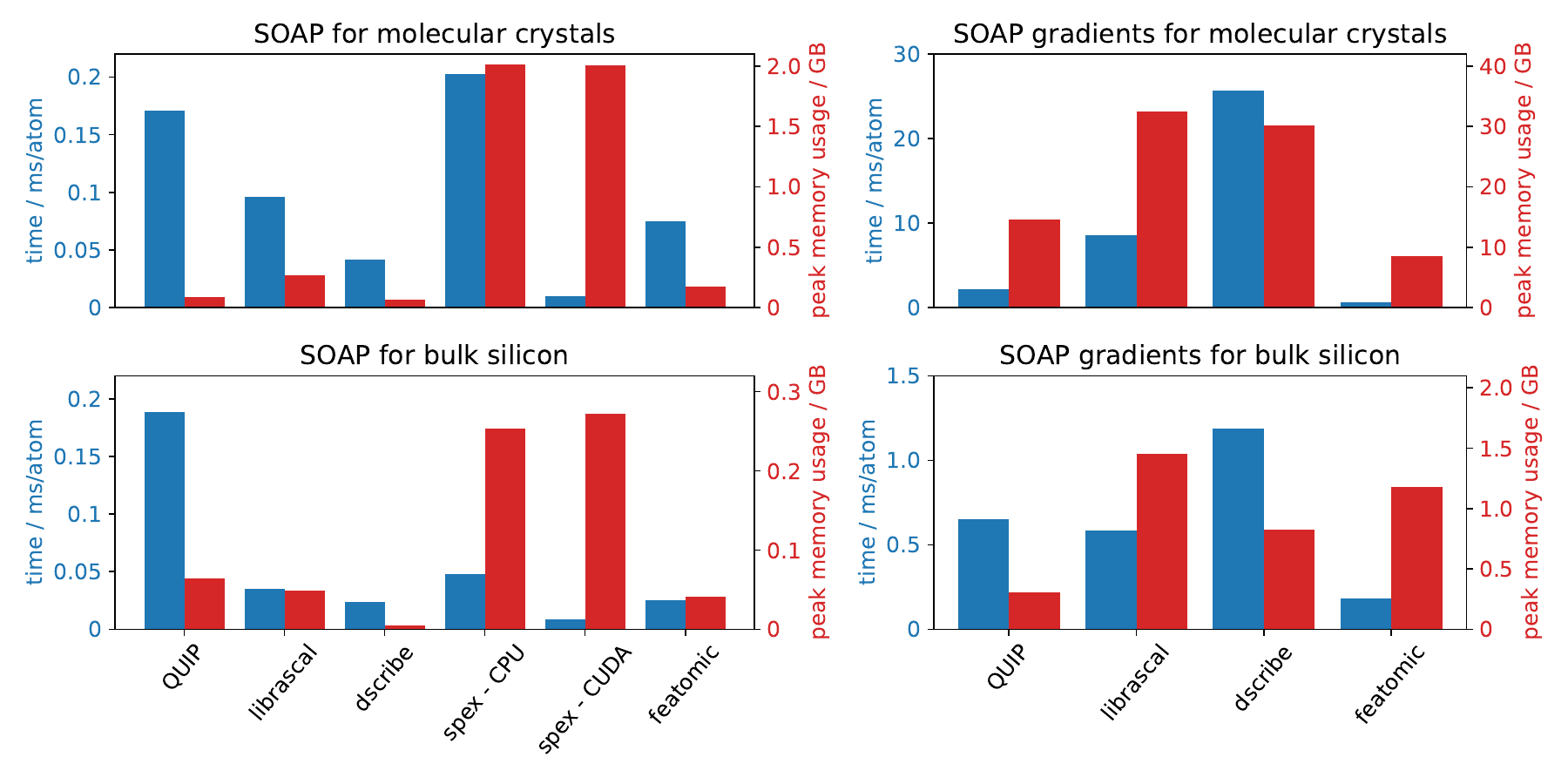}
    \vskip-1.5em
    \caption{Performance and memory efficiency of the \texttt{featomic} and \texttt{torch-spex} libraries compared to other libraries for the calculation of SOAP power spectrum representation, and the gradient of the representation with respect to atomic positions. We recorded both the average time per atom and the peak memory usage for two datasets: one containing molecular crystals with up to 188 atoms (top), and one containing bulk silicon structures with up to 54 atoms (bottom).
    The CPU timings were obtained on an AMD 5945WX using up to 12 threads, while the CUDA timings were obtained on an NVIDIA H100. The CUDA benchmark also reports GPU memory usage instead of CPU memory usage. No gradient calculation was performed for \texttt{torch-spex} as computing gradients of the representation vector through PyTorch's autograd was found to be prohibitively expensive. \resub{High memory usage of \texttt{torch-spex} is due to PyTorch overhead and \texttt{torch-spex} not being designed for sparse representations, while the SOAP power spectrum is highly sparse.}
  }
  \label{fig:featomic}
  \vskip-1em
\end{figure*}

\texttt{featomic} is a library to compute descriptors (sometimes also referred to as representations) for atomistic systems. Its core functionality is implemented in Rust, and interfaces are also available to C, \cxx and Python through a common C API, similar to \metatensor. The Python package also contains utilities for the evaluation of more complex representations and manipulation of spherical tensors.

Compared to existing descriptor libraries, such as \texttt{QUIP}\cite{Bartok2010}, \texttt{DScribe}\cite{Himanen2020} and \texttt{librascal}\cite{musi+21jcp} (from which it is inspired), it offers a unique set of advantages, including: parallelism in a shared-memory context; integration with the automatic differentiation framework in \texttt{PyTorch} through \texttt{featomic-torch}; and memory efficiency. The timings comparison for the evaluation of the SOAP power spectrum descriptor is shown in Figure~\ref{fig:featomic}, along with the peak memory usage for the computation. One can see that \texttt{featomic} is among the fastest implementations when computing the descriptors, and consistently faster when computing gradients of the descriptors. It also uses a smaller amount of memory, especially when computing gradients where it can take advantage of the sparsity of the data when storing it with \metatensor. For example, computing the gradients for the molecular crystals dataset requires a peak memory usage of 8GiB for \texttt{featomic}, compared to around 30GiB for both \texttt{librascal} and \texttt{DScribe}, and 15GiB for \texttt{QUIP}.

Multiple representations are implemented in \texttt{featomic}, including spherical expansions of the atomic density, which are the basis of the SOAP\cite{bart+13prb} and ACE\cite{drau19prb} families of representations; a local expansion of long-range densities called LODE\cite{loch+25jcp, gris-ceri19jcp} which can be used to incorporate long-range effects in ML models; a pair-decomposed version of the spherical expansion, used to generate multi-center representations\cite{niga+22jcp, niga+22jcp2}, as well as a handful of others. \resub{For each representation, we also provide a manual analytical implementation of gradients with respect to atomic positions, strain, and unit cell matrix.}

\texttt{featomic} also includes a set of Python utilities for performing operations on \texttt{TensorMap} objects representing data with O(3) symmetry. These include Python calculators that generate higher-body-order equivariant descriptors from low-body-order density expansion using Clebsch-Gordan tensor products\cite{niga+20jcp}. For example, \texttt{featomic} provides a calculator to compute the equivariant power spectrum (otherwise known as a $\lambda$-SOAP, a representation with body order three\cite{gris+18prl}) by combining two spherical expansions of the atomic density.

\subsection{torch-spex}

\texttt{torch-spex} is a library based on \texttt{PyTorch} and \texttt{sphericart} that computes spherical expansions, i.e., features describing local atomic neighborhoods~\cite{musi+21cr}. It also provides the building blocks of such representations, so end users can design their own featurizations or use these building blocks to construct equivariant message passing neural networks. The main difference with \texttt{featomic} is that \texttt{torch-spex} supports GPU acceleration and automatic differentiation by virtue of being based on \texttt{PyTorch}. \texttt{torch-spex} offers both a pure \texttt{PyTorch} interface as well as a \metatensor interface which returns \texttt{TensorMap}s that are compatible with the output of \texttt{featomic} (Section~\ref{sec:featomic}).

\begin{figure}[htbp]
  \centering
  \begin{tikzpicture}[x=1cm,y=1cm]

\node () at (-1.65,0) {\large $\sum_j {\mathbf C}(Z_j) {\mathbf  R}^l(r_{ij}) {\mathbf S}^{l}(\vec r_{ij})$} ;
\draw[draw=black,fill=red,fill opacity=0.5] (-3.1,-0.25) rectangle ++(1.13,0.6);
\draw[draw=black,fill=orange,fill opacity=0.5] (-3.1+1.13,-0.25) rectangle ++(1.23,0.6);
\draw[draw=black,fill=green,fill opacity=0.5] (-3.1+1.13+1.23,-0.25) rectangle ++(1.15,0.6);

\draw[->,thick] (0.1,0.5) -- (1.0,1.2) ;
\node () at (1.5,2.8) {\includegraphics[scale=0.11]{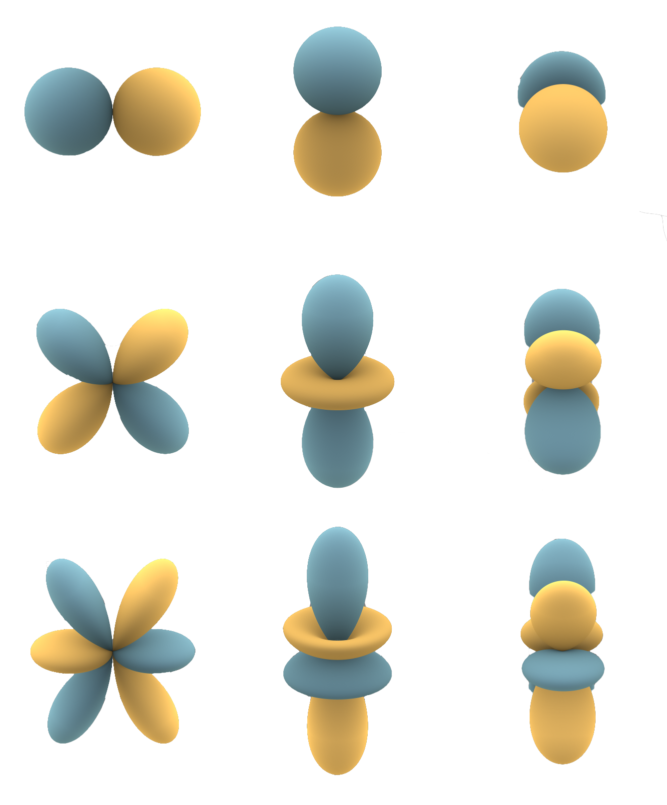}};

\draw[->,thick] (-2.5,0.5) -- (-2.5,1.2) ;

\node () at (-4.4,4.3+0.2) {\textbf{Orthogonal}} ;
\node () at (-2.85,3.5+0.2) {$\text{H} = \begin{bmatrix}
    1.0 \\
    0.0 \\
    0.0
    \end{bmatrix}$\vspace{0.5cm} $\text{C} = \begin{bmatrix}
    0.0 \\
    1.0 \\
    0.0
    \end{bmatrix}$ \vspace{0.5cm} $\text{O} = \begin{bmatrix}
    0.0 \\
    0.0 \\
    1.0
    \end{bmatrix}$};

\node () at (-4.4,2.55+0.2) {\textbf{Alchemical}} ;
\node () at (-2.85,1.7+0.2) {$\text{H} = \begin{bmatrix}
    0.3 \\
    0.6 \\
    0.1
    \end{bmatrix}$\vspace{0.5cm} $\text{C} = \begin{bmatrix}
    0.7 \\
    0.1 \\
    0.2
    \end{bmatrix}$ \vspace{0.5cm} $\text{O} = \begin{bmatrix}
    0.3 \\
    0.3 \\
    0.4
    \end{bmatrix}$};

\draw[->,thick] (-1.4,-0.35) -- (-1.4,-1.1) ;

\node () at (0.8,-2.2) {\includegraphics[scale=0.35]{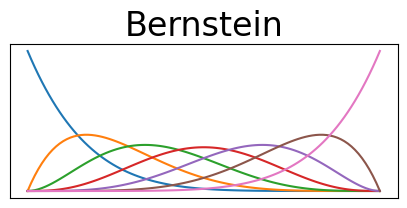}};
\node () at (-1.35,-4.2) {\includegraphics[scale=0.35]{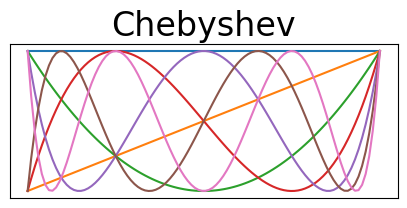}};
\node () at (-3.5,-2.2) {\includegraphics[scale=0.35]{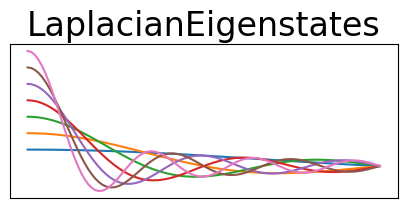}};

\end{tikzpicture}
  \caption{Illustration of a spherical expansion: The chemical species $Z_j$, the neighbor distance $r_{ij}$, and the direction $\vec r_{ij}$ can be expanded in different ways.}
  \label{fig:spex}
\end{figure}

This flexibility and scalability are essential when designing neural networks; the various options for the computation of descriptors in \texttt{torch-spex} are shown in Table~\ref{tab:torch-spex}. Different radial bases (i.e., expansions of interatomic distances), angular bases (i.e., ways to expand orientation), and chemical embeddings (i.e., ways to model different atomic species), are all supported. Both dense chemical embeddings, where atomic species are mapped into a reduced space of ``pseudo species''\cite{willatt2018feature}, and sparse chemical embeddings --- where atomic species are kept in separate sectors of the descriptor array --- are supported.

\begin{table}
  \caption{List of descriptor options in \texttt{torch-spex}.}
  \setlength{\tabcolsep}{10pt}
  \begin{tabular}{l|c}
    \toprule
    Basis & Options \\
    \midrule
    \multirow{2}{*}{Angular} & spherical harmonics, \\
     & solid harmonics \\
     \midrule
    \multirow{2}{*}{Radial}  & spherical Bessel,\\ & Bernstein, Chebyshev \\
    \midrule
    \multirow{2}{*}{Chemical} & orthogonal (sparse),\\
     & alchemical (dense embedding) \\
    \bottomrule
  \end{tabular}
\label{tab:torch-spex}
\end{table}

\subsection{torch-pme}

\texttt{torch-pme}\cite{loch+25jcp} is a PyTorch-based library for the efficient computation of long-range atomistic interactions with automatic differentiation support. It provides a complete suite for long-range calculations, including Particle-Particle Particle-Mesh Ewald (P3M), Particle Mesh Ewald (PME), standard Ewald, and non-periodic calculations. The library can compute any integer exponent $1/r^p$ potential, enabling various long-range interactions such as charge-dipole, dipole-dipole, dispersion, and magnetostatics. This flexibility allows cutting-edge ML potentials to incorporate long-range interactions for large-scale systems. Optimized for both CPU and GPU devices, \texttt{torch-pme} is fully compatible with TorchScript for Python-independent execution in high-performance simulation engines. The library provides a \metatensor API, storing outputs as \texttt{TensorMap}s and accepting \metatomic inputs, to enable seamless integration of long-range interactions in existing ML workflows. \resub{An experimental JAX version, \texttt{jax-pme}, is under development and will be integrated with \metatomic once it supports JAX models.}

\begin{figure}[htbp]
  \centering
  \includegraphics{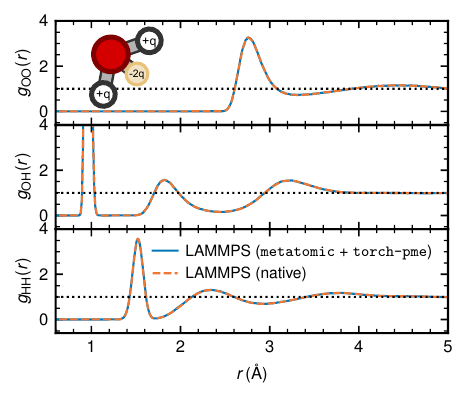}
  \caption{\label{fig:water-rdf}
  Oxygen-oxygen, oxygen-hydrogen, hydrogen-hydrogen radial distribution functions $g(r)$ of the q-TIP4P/F quantum water model from NVT simulations at 300~K using the \metatomic and LAMMPS interface.}
\end{figure}

As an example, we present an implementation of the q-TIP4P/F\cite{habershon_competing_2009} quantum water model as a \metatomic model. The implementation includes bond and angle interactions, non-bonded Lennard-Jones interactions, and Coulomb interactions handled by \texttt{torch-pme}. PyTorch's automatic differentiation capabilities simplify the computation of forces acting on the 4th point, which traditionally requires additional analytic terms.
We run a system of 216 water molecules in a cubic box of 35 \AA{} at 300~K in the NVT ensemble\cite{bussi_canonical_2007}, using a time step of 0.5~fs and a simulation length of 15~ns. The system is initialized with random velocities and equilibrated for 100 ps before collecting statistics for the radial distribution function (RDF). The RDFs are computed using the MDAnalysis\cite{michaudagrawal_mdanalysis:_2011} package. The simulation is run on a single core and one GPU using the non-KOKKOS version of the \metatomic LAMMPS. See section \ref{sec:lammps} for details. The \metatomic implementation of this traditional potential is more flexible but almost 12 times slower compared to the native version in LAMMPS, \resub{which is to be expected when comparing interpreted TorchScript to a highly optimized \cxx implementation}. Figure \ref{fig:water-rdf} compares the resulting RDFs from the \metatomic model with a LAMMPS implementation, showing perfect agreement.

\subsection{vesin}

\texttt{vesin} is a small library that finds all pairs of atoms whose distance is smaller than some cutoff distance (also called neighbor lists) using the cell lists algorithm\cite{alle-tild17book} to obtain $\mathcal{O}(N)$ scaling.

Implemented in \cxx, it offers C, \cxx, Fortran, Python and Torch (both PyTorch and TorchScript) APIs. \texttt{vesin} also focuses on being small and easy to embed in pre-existing software, providing a single file, amalgamated build for integration in \cxx project regardless of the build system. This makes it useful when integrating \metatomic models in simulation engines that do not already provide facilities to compute neighbor lists. We used \texttt{vesin} in the ASE, PLUMED, and eON interfaces described in this work.

\subsection{sphericart}

\texttt{sphericart} is a library for the fast evaluation of spherical harmonics using the algorithm presented in Ref.~\citenum{bigi+23jcp}. It is implemented in \cxx and CUDA, and it exposes interfaces for C, \cxx, Python (using NumPy arrays), PyTorch, and JAX. A native Julia implementation of the same algorithm is also available. \texttt{sphericart} puts an emphasis on obtaining real spherical harmonics directly from Cartesian coordinates, as this is generally the ideal format for spherical harmonics in the context of atomistic applications. Furthermore, sphericart focuses on performance and parallel scaling, providing the efficient calculation of spherical (and solid) harmonics and their derivatives up to second order. Sphericart is developed following a multi-language and multi-paradigm philosophy, allowing, for example, the integration of sphericart's derivatives with automatic differentiation engines such as torch and JAX.

The \metatensor interface of \texttt{sphericart} allows users to obtain the spherical harmonics as a \texttt{TensorMap}, which is a particularly suitable format to store the data (and metadata) of spherical harmonics of different degrees.

\section{An ecosystem of machine-learning models}

Multiple machine learning models have been built using the tools we presented so far. Here we showcase three that address very different learning targets and showcase the flexibility of this software ecosystem.

\subsection{PET-MAD}
\label{sec:pet-mad}

PET-MAD is a universal interatomic potential designed for running complex atomistic simulation workflows across the periodic table for both organic and inorganic materials\cite{mazitov2025petmad}. It is based on a robust and flexible Point-Edge Transformer (PET) architecture\cite{pozd-ceri23nips} and trained on the Massive Atomic Diversity (MAD)\cite{mazitov2025massive} dataset, which incorporates a high degree of chemical and structural diversity while employing highly converged, internally consistent reference calculations. Using fewer than 100,000 structures spanning 85 elements, PET-MAD provides a competitive level of accuracy while being fast and enabling high generalization without sacrificing computational efficiency.

PET-MAD achieves competitive or superior accuracy\cite{mazitov2025petmad} compared to other widely used universal MLIPs in benchmarking, particularly excelling in molecular systems and low-dimensional materials, and rivaling the bespoke system-specific models even in complex settings like ionic transport, phase transitions, surface segregation, while being applicable also for NMR crystallography and capturing quantum nuclear effects. Its high computational efficiency, unconstrained architecture, and integrated uncertainty quantification make it ideal for both exploratory simulations and high-precision workflows.

\begin{figure}
  \centering
  \includegraphics[width=\linewidth]{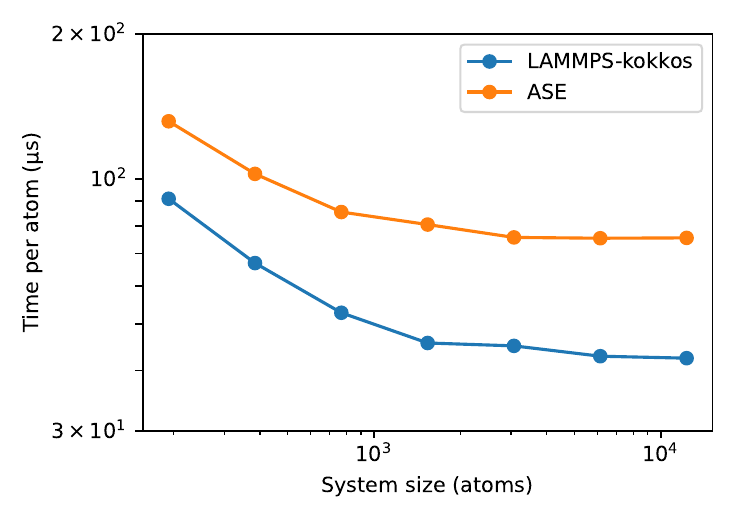}
  \caption{Performance of the PET-MAD model in LAMMPS (with KOKKOS support) and ASE on liquid water system. The evaluation is done in a single NVIDIA H100 GPU. LAMMPS interface of the \metatomic package provides a significant speedup compared to the ASE due to a full GPU support.}
  \label{fig:PET-MAD-lammps-ase}
\end{figure}

One of the key features of PET-MAD is its integration in the \metatensor ecosystem: we employed the \metatrain package for scalable training, evaluation, fine-tuning, transfer learning, and uncertainty quantification within the last-layer prediction rigidity approach\cite{chon+25fd}. This integration enables robust training workflows, making it possible to refine PET-MAD for specific applications if needed, requiring minimal data and computational overhead, while maintaining performance on diverse systems. After this fine-tuning, PET-MAD can be exported to a \metatomic model, facilitating seamless application with major simulation engines. This pipeline supports direct use in production environments and advanced simulations, including replica exchange molecular dynamics, and path integral simulations and advanced sampling. \resub{Figure~\ref{fig:PET-MAD-lammps-ase} demonstrates the performance of the PET-MAD in the ASE \cite{larsen2017atomic} and LAMMPS \cite{thompson2022lammps} engines, where the latter provides a significant speedup thanks to a full GPU implementation (see more details in Section~\ref{sec:lammps} and \ref{sec:ase}). A systematic study of the PET-MAD accuracy and performance, as well as a comparison against other existing MLIPs can be found in the original article, reference~\citenum{mazitov2025petmad}.} Therefore, through its close alignment with the \metatensor infrastructure, PET-MAD not only offers accuracy but also practical usability, making it a cornerstone model in the emerging ecosystem of interoperable, transferable tools for atomistic simulations.

\subsection{ShiftML}

\begin{figure}[htbp]
  \centering
  \includegraphics[width=\linewidth]{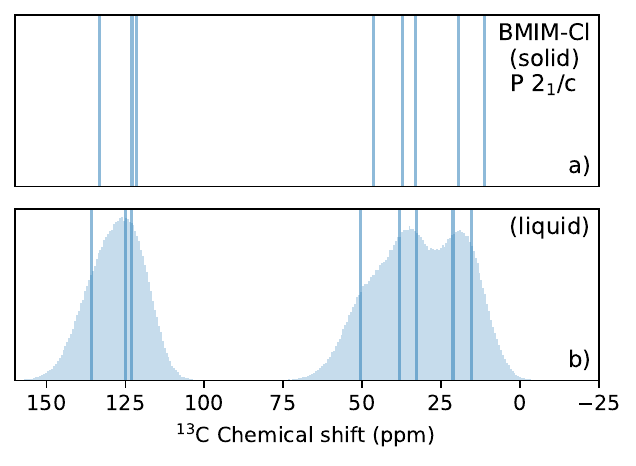}
  \caption{Chemical shifts of BMIM-Cl solid (P21/c) (a) and liquid BMIM-Cl (b).
  Vertical lines denote chemical shifts computed with ShiftML3 for the PET-MAD geometry relaxed structure (a)  and averages from a PIMD simulation discussed in Section~\ref{sec:i-pi} (b).
  Furthermore, we show the overall histogram  of chemical shifts in the PIMD trajectory in panel (b).}
  \label{fig:shiftml}
\end{figure}

ShiftML is an umbrella name for ML models that can predict chemical shieldings in organic crystals, which have been developed and improved since 2018\cite{paru+18ncomm, enge+19pccp, cord+22jpcc, kellnerDeepLearningModel2025a}. Fast predictions of chemical shieldings are essential for the correlation of atomistic models and nuclear magnetic resonance (NMR) spectroscopy experiments. In a procedure dubbed NMR crystallography, structural candidates are ranked according to the likelihood of predicted chemical shifts matching experimentally measured values\cite{emsleySpiersMemorialLecture2025a, hodgkinsonNMRCrystallographyMolecular2020b, harrisNMRCrystallography2009}. \resub{Both experimentalists and computational scientists employ computational shielding models on a regular basis to determine the structure of organic solids.\cite{rehmanOrganicNMRCrystallography2025, l.guestEssentialSynergyMD2025,cordovaStructureDeterminationAmorphous2021a}}

Initially, ShiftML was based on SOAP Kernel models\cite{paru+18ncomm, enge+19pccp, cord+22jpcc} and was limited to predict shieldings in organic crystals with at most four elements (H,C,N,O)\cite{paru+18ncomm}. More recently, the domain of applicability of ShiftML has been extended to organic crystals with up to 12 species and thermally distorted geometries by creating larger and more diverse training sets (ShiftML2)\cite{cord+22jpcc}.

Including more chemical species and diverse chemical structures increases both the computational demand of model training and increases the \resub{SOAP power spectrum} size drastically due to its quadratic scaling with the number of unique chemical elements in the dataset, \resub{a scaling that can be partially mitigated by either using some form of feature contractions, or a block sparse storage format for descriptors such as \metatensor}.\cite{de+16pccp, lopa+23prm, darbyCompressingLocalAtomic2022, niga+20jcp}

\resub{We found that modern atomistic deep learning architectures overall make more accurate chemical shielding predictions than SOAP-based models, and did not require any special handling to scale to many chemical elements.} Therefore, in the latest release of ShiftML (ShiftML3), we modernized the infrastructure of ShiftML and moved it to a modern deep learning architecture. ShiftML3 was created as a committee of PET models\cite{pozd-ceri23nips} using \metatrain and exported to a \metatomic model. The ShiftML Python package then uses the ASE compatibility layer of \metatomic to facilitate easy integration of the ShiftML model into various NMR crystallographic workflows. This package hosts the latest ShiftML3.0 model and is intended to host future development versions of ShiftML.

In Figure~\ref{fig:shiftml} we compute chemical shieldings of crystalline BMIM-Cl (Spacegroup P2$_{1}$/c\cite{holbreyCrystalPolymorphism1butyl3methylimidazolium2003}) as well as chemical shieldings of liquid BMIM-Cl. For the computation of the crystalline BMIM-Cl, we first relax the positions of the experimental crystal structure and then compute chemical shieldings through the ASE integration of ShiftML3 and PET-MAD (see also Section~\ref{sec:pet-mad}  \resub{or additional details of the simulation parameters}). To compute chemical shieldings of liquid BMIM-Cl, we perform Path Integral Molecular Dynamics simulations of a box of 16 ion pairs \resub{at 500~K} employing the PET-MAD potential through its \metatomic integration in the i-PI molecular dynamics simulation engine (see Section~\ref{sec:i-pi}). We then compute chemical shielding averages from the molecular dynamics simulation using ShiftML3 and its ASE integration. The combined simulation of molecular dynamics sampling, considering explicit quantum nuclear effects, followed by the post-hoc computation of an experimental observable, demonstrates the seamless interoperability of \metatensor components for accurate property predictions of molecular solids and liquids.

\subsection{FlashMD}
\label{sec:flashmd}

\begin{figure}[htbp]
  \centering
  \includegraphics[width=\linewidth]{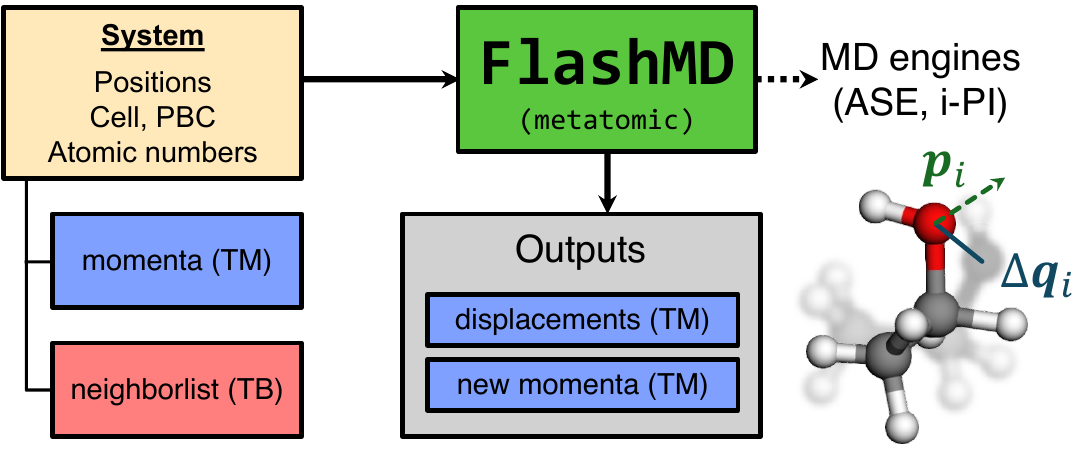}
  \caption{A schema explaining the utilization of \metatrain and \metatomic in the FlashMD workflow. TM stands for \texttt{TensorMap}, and TB stands for \texttt{TensorBlock}. Along with the conventional \texttt{System} object, FlashMD presently accepts the momenta and neighbor lists as additional inputs to predict the displacements ($\Delta \mathbf{q}_i$) and the updated momenta ($\mathbf{p}_i$) after a large time step.}
  \label{fig:flash-MD}
\end{figure}

FlashMD\cite{bigi2025flashmd} is a machine-learning method that directly predicts molecular dynamics trajectories. Compared to traditional molecular dynamics using machine-learned interatomic potentials, FlashMD bypasses the costly evaluation of forces via gradients of the potential energy surface and directly predicts the future positions and momenta. Models are trained to predict time steps much larger than conventional molecular dynamics, which allows practitioners to dramatically reduce the number of model evaluations and accelerate molecular dynamics by one or two orders of magnitude~\cite{bigi2025flashmd}. \metatrain and \metatomic were used to train and distribute universal models for the prediction of molecular dynamics trajectories across the periodic table, and to achieve their integration into the ASE and i-PI molecular dynamics engines. The development of FlashMD showcases the flexibility and generality of the \metatensor ecosystem. Figure~\ref{fig:flash-MD} shows how \metatensor and \metatomic data classes can seamlessly represent inputs and outputs of FlashMD models, allowing for their integration with external molecular simulation codes.

\section{Integrations with other simulation and analysis tools}
\label{sec:engines}

\subsection{LAMMPS}
\label{sec:lammps}

\begin{figure}[htbp]
  \centering
  \includegraphics[width=1.0\linewidth]{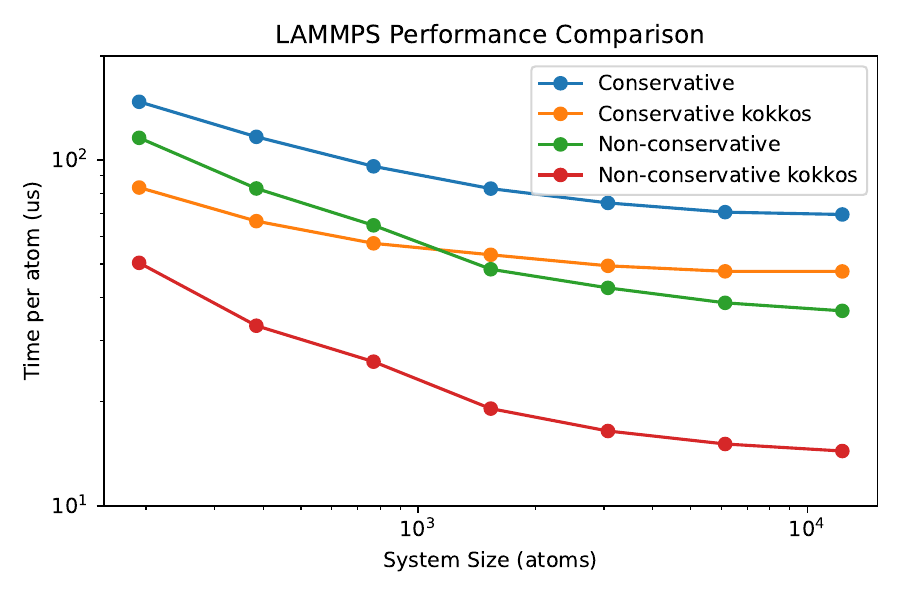}
  \caption{Performance comparison of the metatomic interface, in KOKKOS and non-KOKKOS mode, and running conservative and non-conservative molecular dynamics, using the PET-MAD universal potential for a liquid water system.}
  \label{fig:lammps}
\end{figure}

LAMMPS (Large-scale Atomic/Molecular Massively Parallel Simulator)\cite{thompson2022lammps} is an open-source molecular dynamics simulator. It is highly flexible and scalable, making it suitable for the atomic-scale simulation of materials, as well as biological and soft matter systems,  on parallel computing architectures. The integration of \metatomic models in LAMMPS -- similar to all molecular dynamics engines mentioned below -- is based on the \metatomic model providing energies, forces, and stresses to the simulation engine at every discrete time step, given the particle types and positions (and cell matrix, for periodic systems) from the simulation.

Two variations of the \metatomic interface to LAMMPS are available. The first is a traditional interface where the simulation itself runs on CPU, and the model can run on CPUs and GPUs through PyTorch's ability to execute the same TorchScript model on multiple backends. This first interface requires host-to-device data transfers (and vice versa) at each timestep of the simulation which is limited in speed. The second interface is based on the KOKKOS-accelerated version of LAMMPS, where all atomic data is stored on the GPU (or any other accelerator), avoiding any data transfer between the CPU and the GPU. The performance of the traditional and KOKKOS interfaces are compared in Figure~\ref{fig:lammps} for simulations of liquid water boxes of different sizes. The figure also compares the cost of simulations using forces computed as a direct output of the model -- which is faster, but introduces systematic sampling errors, because they do not ensure energy conservation\cite{bigi+25icml} -- and those computed, as customary, as the derivatives of the potential. \metatomic models can be trained to evaluate both types of outputs, and the LAMMPS interface further provides the functionality to run multiple-time-step simulations that offer the acceleration of direct force predictions while ensuring energy conservation and correct sampling.
More details are given in the discussion of the integration with i-PI, Section~\ref{sec:i-pi}.

\subsection{i-PI}
\label{sec:i-pi}

i-PI is a Python code, originally developed\cite{ceri+14cpc} to perform path integral molecular dynamics, an advanced sampling technique that provides a quantum mechanical description of the nuclear degrees of freedom (so-called nuclear quantum effects)\cite{mark-ceri18nrc}. It is based on a client/server model, in which energy, forces, stress -- that are needed to evolve the atomic coordinates in time and sample the desired thermodynamic ensemble -- are evaluated by any external code supporting its minimalist communication protocol. More recently, i-PI has been extended to incorporate several advanced sampling techniques and has been made sufficiently fast to perform simulations with efficient, highly-parallelized ML potentials\cite{litm+24jcp}. Its flexible, modular structure is ideal to exploit the advanced ML functionalities of \metatomic models, also thanks to a dedicated driver that can be used both as a library and as a stand-alone program.

As a demonstration, we use it to perform a path integral simulation of the ionic liquid BMIM-Cl, computing the mean potential, and the centroid-virial kinetic energy estimator for different atomic species. We use a ML model (PET-MAD, also discussed in Section~\ref{sec:pet-mad}) that computes interatomic forces both as derivatives of the potential, and as non-conservative forces. The latter approach is faster, but leads to forces that violate energy conservation\cite{bigi+25icml}, and to sampling errors that can be reduced, but not eliminated, by the use of efficient thermostating schemes. Fortunately, it is possible to recover most of the speed-up of non-conservative forces by using (i) a multiple-time-stepping algorithm\cite{tuck+92jcp} in which the conservative forces are computed every few steps as a correction, and (ii) a ring-polymer contraction scheme\cite{mark-mano08cpl} that performs a similar operation along the beads of a ring polymer. The two methods can often be used in tandem\cite{kapi+16jcp}, as we do here. We use the global version of the path-integral Langevin equation thermostat\cite{ceri+10jcp}, which controls the temperature of the internal degrees of freedom of the ring polymers with optimal coupling strength, and uses a stochastic velocity rescaling thermostat\cite{buss+07jcp} for the centroid. We run simulations with 32 beads and a time step of 0.5~fs, at a temperature of 500~K.

\begin{figure}[htbp]
  \centering
  \includegraphics[width=1.0\linewidth]{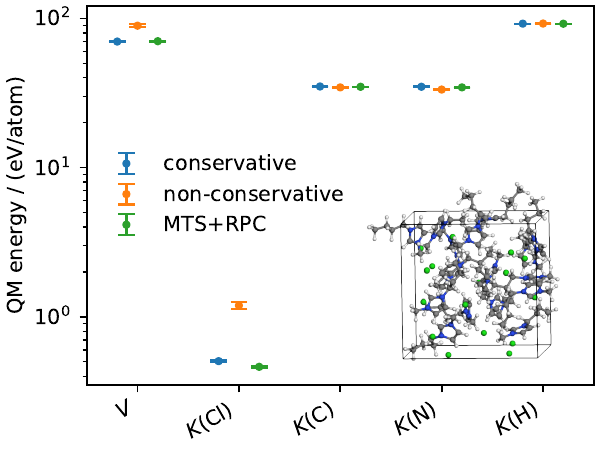}
  \caption{Mean quantum potential and quantum kinetic energies for a simulation of 1-Butyl-3-methylimidazolium chloride at 500~K, computed by subtracting from the path integral estimators the averages computed from a conservative classical simulation.}
  \label{fig:ipi}
\end{figure}

Using these non-conservative forces leads to a large drift of the conserved quantity of the simulation, and a deviation of the potential energy from the reference of a conservative simulation, see Figure~\ref{fig:ipi}. However, the quantum kinetic energy of different species shows only small deviations from the conservative target, which indicates that the strong local thermostatting of the ring polymer is able to compensate for the lack of energy conservation: the largest (relative) error is observed for Cl atoms, that are heavier and behave almost as classical particles. Using a ring-polymer contraction to 4 beads and a multiple time-stepping factor of 8 (which reduces by a factor of 64 the number of conservative force evaluations, recovering almost entirely the speed-up of direct-force evaluation) eliminates these small errors, and yields results for the quantum thermodynamic average energetics in quantitative agreement with the conservative simulations.

\subsection{eOn}

\texttt{eOn} is a software package with algorithms geared towards the exploration of potential energy surfaces as a function of their critical points. These critical points are calculated using state of the art saddle search methods\cite{asgeirssonNudgedElasticBand2021,goswamiEfficientImplementationGaussian2025a} in a \cxx client which can call a \metatomic model for the potential energy and forces. The core state object in \texttt{eOn} does not include neighbor lists, so the interaction to \metatomic is in turn facilitated by using \texttt{vesin}. Several trial saddle point configurations are generated from a \texttt{Python} server, which successively builds up information on the current configuration. When the escape routes from a given state are computed up-to a tolerance, the server takes a Kinetic Monte Carlo step, typically under the Harmonic Transition State theory assumption, thus resulting in an off-lattice, or ``adaptive'' Kinetic Monte Carlo\cite{henkelmanLongTimeScale2001,trochetOfflatticeKineticMonte2018}. The emphasis is on the long time scale evolution of atomic systems, where the dynamics of interest can be described by fast transitions between stable states. Coupling to \metatomic models enables the study of large systems with high accuracy over a range of temperatures and times not typically accessible to explicit electronic structure calculations. Adaptive kinetic Monte Carlo (aKMC)\cite{pedersenDistributedImplementationAdaptive2010} \resub{generates an "on-the-fly" catalog from critical points for off-lattice kinetic Monte Carlo. This approach allows for the study of complex systems, such as catalytic events on surfaces and surface ripening. Such phenomena are often intractable for conventional methods. Empirical force fields, for example, may prove too jagged or fail to capture the physics of interest. Higher-order calculations, in turn, are constrained by prohibitive computational scaling, a limitation that restricts their application to small systems which can introduce finite-size effects.} Thus, integration with \metatomic addresses one of the key challenges, namely, the cost to compute samples for saddle searches of states\cite{henkelmanAtomisticSimulationsActivated2017,henkelmanLongTimescaleSimulationsChallenges2018} in the wider applicability of such off-lattice KMC methods to the study of activated processes.

\begin{figure}[htbp]
  \centering
  \includegraphics[width=1.0\linewidth]{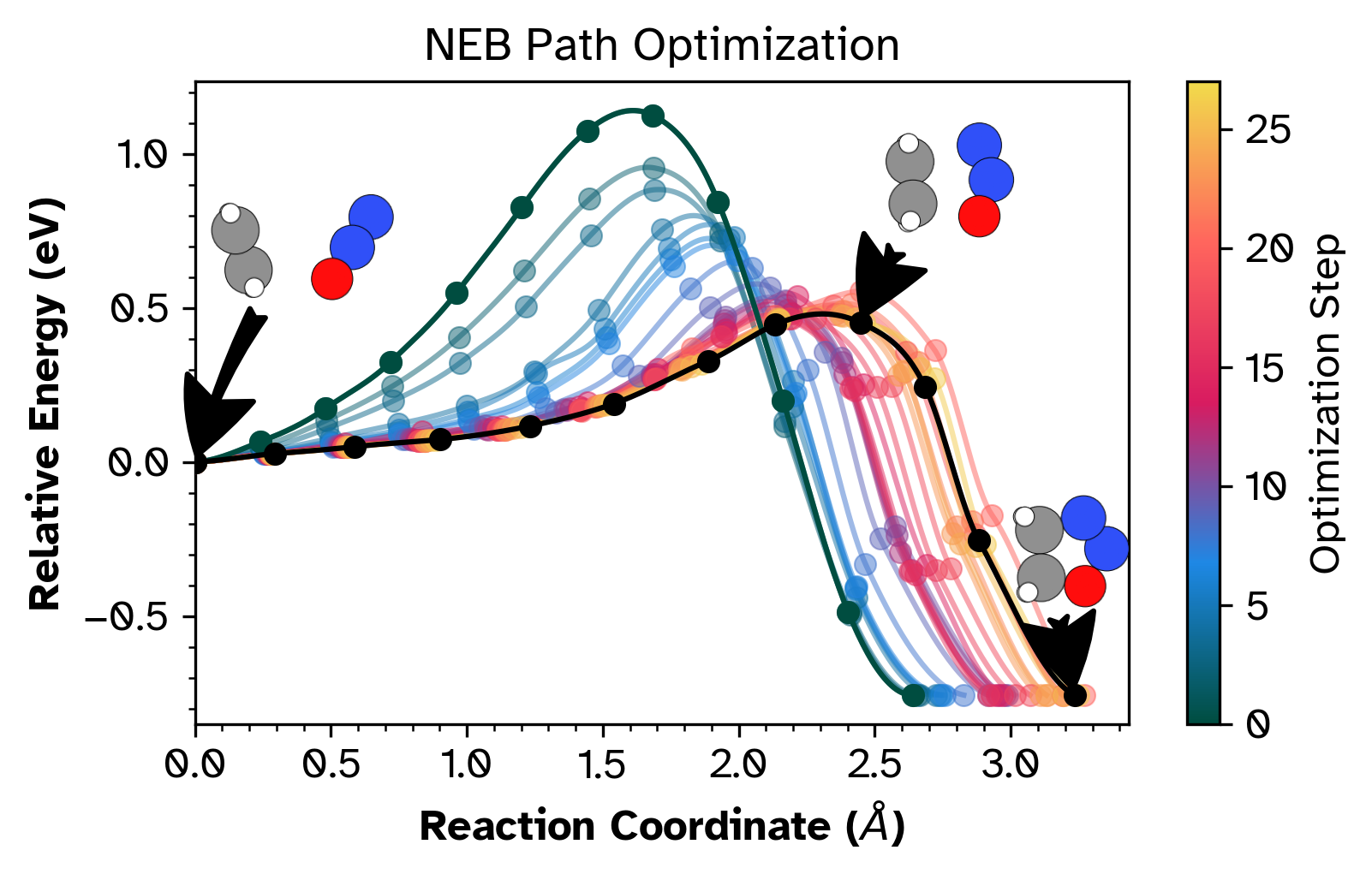}
  \caption{Progression of configurations at each step of the Nudged Elastic Band calculation using the PET-MAD energy surface. \resub{The teal line indicates the initial path generated using the IDPP method\cite{smidstrupImprovedInitialGuess2014} after permutation matching of the endpoints with the IRA method \cite{gundeIRAShapeMatching2021}}. The black line and inset images indicate the final converged path, the convergence of intermediate paths is indicated by the colorbar \resub{which demonstrates standard relative energy deviations for NEB calculations.}Reaction coordinate is taken as the total path length, calculated as the sum of the distances between images on the path.}
  \label{fig:eon_ewneb}
\end{figure}

As a demonstration, we will consider the reaction of ethylene and N$_2$O to form 4,5-Dihydro-1,2,3-oxadiazole with ten intermediate images\cite{koistinenMinimumModeSaddle2020} using the PET-MAD potential energy surface without fine tuning. Resolving the transition state with such a low number of images is difficult with standard NEB calculations. In Figure~\ref{fig:eon_ewneb} we show that starting from an image-dependent path potential\cite{smidstrupImprovedInitialGuess2014} and using energy-weighted strings\cite{asgeirssonNudgedElasticBand2021} ensures fast and efficient convergence. This is accelerated by intermittently taking up to ten steps of the single-ended dimer minimum-mode-following method from the climbing image along the estimated NEB tangent. The true energy barrier can subsequently be obtained through a few steps of the GP-dimer algorithm\cite{goswamiEfficientImplementationGaussian2025a}, since the geometry of the transition state is correctly identified.

\subsection{TorchSim}

TorchSim\cite{torchsim-arxiv} is a Torch-based atomistic simulation engine, implementing GPU-accelerated integrators and sampling algorithms, and optimized to function together with machine-learning interatomic potentials. It provides an interface to \metatomic models, making it possible to use any compatible model in simulations.

\begin{figure*}
  \centering
  \includegraphics[width=0.8\linewidth]{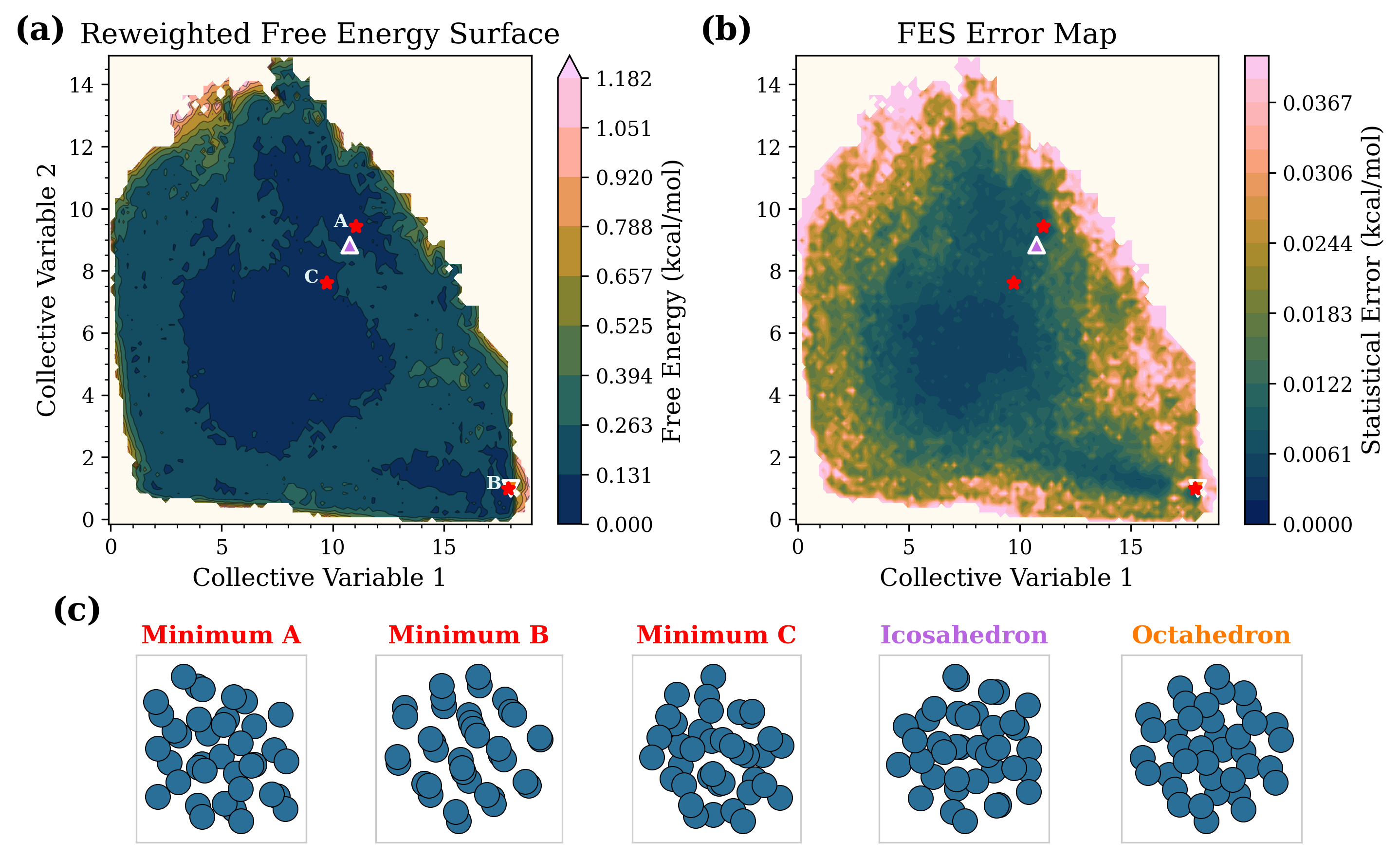}
  \caption{Free energy surface (FES) diagnostics for a Lennard-Jones 38-atom (LJ38) cluster after ten million steps of a well tempered metadynamics run at a temperature of 0.17 in reduced units. We used the integral of coordination number histograms --- implemented as a custom \metatomic model --- as our collective variables. (a) The final FES is estimated via histogram reweighting of the simulation trajectory. (b) The error map is obtained from block averaging over 5000 blocks. The minima are located by grid searches across the bins of the free energy estimate\cite{traplAnalysisResultsMetadynamics2022}. The panel (c) shows known crystalline structures along with the minima found. As expected\cite{ceriottiDemonstratingTransferabilityDescriptive2013} the double funnel is evident, with the distorted icosahedral structures forming one basin, and the global minima of the octahedron forming another.}
  \label{fig:lj38-metad}
  \vskip-1em
\end{figure*}

\subsection{ASE}
\label{sec:ase}

ASE (Atomic Simulation Environment)\cite{larsen2017atomic} is a Python library for setting up, running, and analyzing atomistic simulations. It provides a flexible interface to many quantum chemistry codes and supports scripting of workflows for structure generation, optimization, property calculations, and molecular dynamics simulations.

Thanks to the integration between \metatomic and ASE, \metatomic models can be used to provide energies, forces, and stresses for molecular dynamics simulations and geometry optimization workflows. Within the interface, \texttt{vesin} is used to provide fast neighbor list calculations, which is often a limiting factor for simulations using the native ASE neighbor list calculators. Accelerator devices, such as GPUs, can be used to accelerate model execution, although data transfers between the host and the device will be executed at every step, which is inevitable since ASE's data structures are based on NumPy.

In addition to the energy and its gradients (forces and stress), other model outputs can be used through this interface. This includes the direct prediction of non-conservative forces and stresses, as well as any other custom properties that might be supported by the model but which are not a subset of the standard properties defined by ASE.

\subsection{PLUMED}

\texttt{PLUMED} is a library implementing a collection of methods which facilitate the exploration of the free energy surface of a system as a function of (typically) lower-dimensional variables\cite{tribelloPLUMED2New2014}, colloquially known as ``collective variables'' (CVs). The integration of \metatomic enables users to compute arbitrary CVs with PyTorch code. Within this framework it is possible to efficiently obtain derivatives with respect to structure parameters for features using automatic differentiation and, in principle, allows for the CV to be computed on a GPU or other devices. This --- together with the chemiscope integration described in Section~\ref{sec:chemiscope} --- makes it very easy to try different functional forms and parameters when defining new CVs. In doing so it also makes it possible to use machine learning tools to define CVs, joining the host of custom representations such as the ones computed by \texttt{featomic} or \texttt{torch-spex}; to implicitly learned representations like ATLAS and others\cite{gibertiGlobalFreeenergyLandscapes2021,trizioEnhancedSamplingReaction2021,hanniDataEfficientLearning2025,frohlkingDeepLNELeveragingKnowledge2024,yangLearningCollectiveVariables2024,dietrichMachineLearningNucleation2024,rayDeepLearningCollective2023,spiwokFrontiersCollectiveVariable,bonatiDeepLearningSlow2021,bonatiDatadrivenCollectiveVariables2020,giorginoPYCVPLUMED2,chenMolecularEnhancedSampling2018}.

As a representative application, we demonstrate an exploration of a 38-atom cluster interacting through the Lennard-Jones potential. Despite the apparent simplicity of the potential, the potential energy surface is notoriously complex, featuring a deep global minimum and a broad basin of defective structures\cite{gasparottoRecognizingLocalGlobal2018,ceriottiDemonstratingTransferabilityDescriptive2013,gibertiGlobalFreeenergyLandscapes2021}. An effective collective variable for this system may be based on a B-spline integration of kernel density estimator on the histogram over coordination numbers (CN)\cite{ceriottiDemonstratingTransferabilityDescriptive2013}, which can be implemented efficiently using the neighbor list provided by \texttt{vesin}. Without changes to the PLUMED sources, the CV can be exported as a \metatomic atomistic model for use within PLUMED, which in turn is then called from LAMMPS.

Figure~\ref{fig:lj38-metad} showcases the results of a metadynamics trajectory using the histogram CV implemented \textit{via} \metatomic. The high-dimensional trajectory, described by the CN histograms, is shown as a function of the two collective variables. An estimate of the error is obtained by post-processing the simulation trajectory. Three basins are indicated, with representative configurations of each minimum shown as overlays, thus demonstrating the utility of implementing bespoke, data-driven CVs with \metatomic.

\subsection{chemiscope}
\label{sec:chemiscope}

\texttt{chemiscope} is an interactive structure-property explorer for atomic-scale materials and molecules\cite{frau+20joss}, providing in browser visualization of large databases and helping researchers to find structure—property correlations inside such databases. It was designed with versatility at its core, offering a stand-alone viewer, a Python module, and a widget that can be used inside a Jupyter notebook. Within \texttt{chemiscope}, the \texttt{chemiscope.explore} function provides a streamlined way to visualize datasets by automatically computing a representation and using dimensionality reduction. This allows users to rapidly gain insights into the composition and structure of data without the need to manually implement and fine-tune the representation process.

\begin{figure}[htbp]
  \centering
  \includegraphics[width=\linewidth]{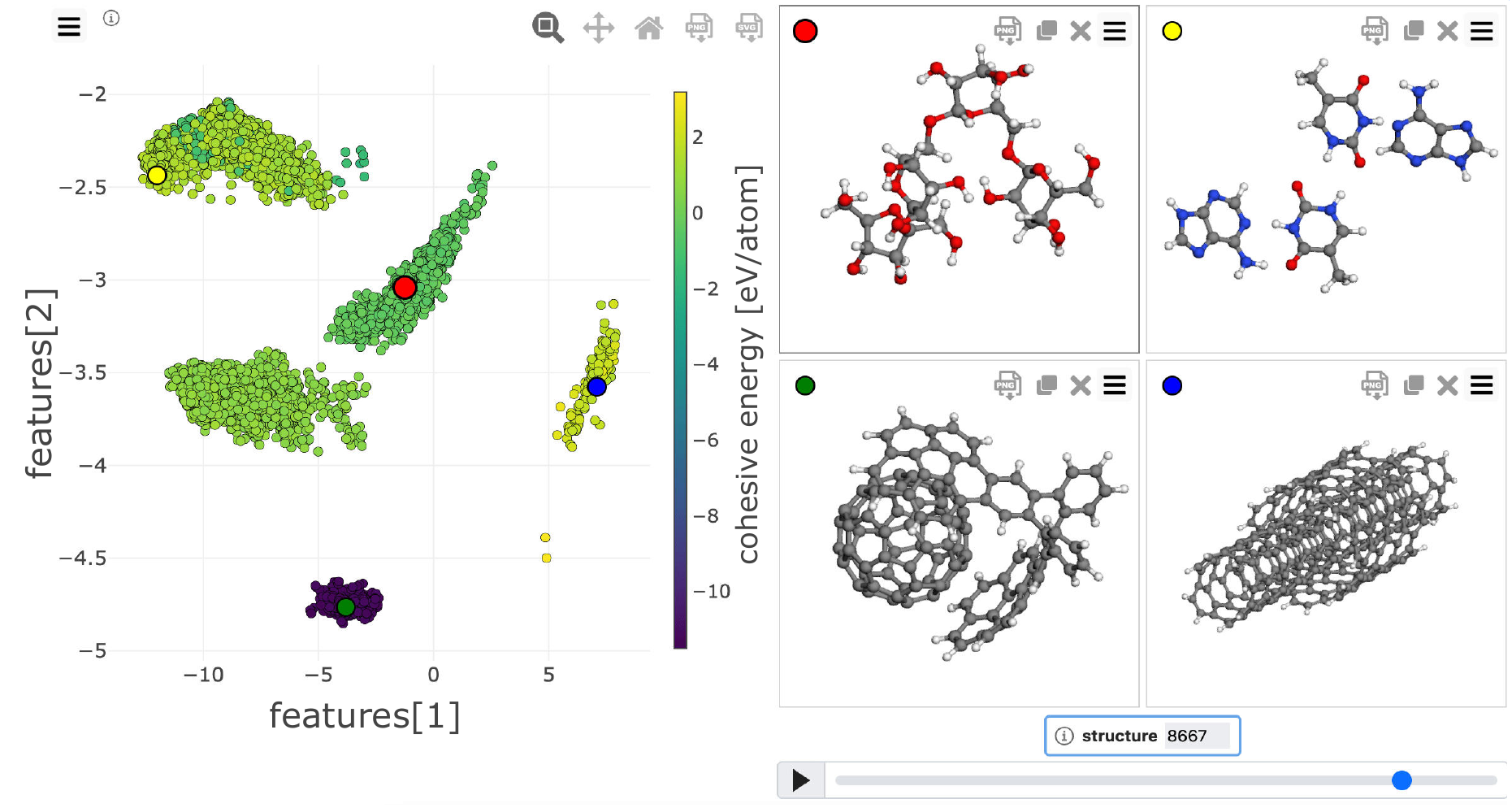}
  \caption{Visualization of the MD22 dataset in chemiscope, using the \texttt{chemiscope.explore} function --- with PET-MAD as a feature generator --- to automatically generate the map on the left-hand side.}
  \label{fig:chemiscope-explore}
\end{figure}

In addition to pre-defined descriptor calculators, \texttt{chemiscope.explore} accepts \metatomic model instances to compute a featurization of the systems, making it possible for users to employ the latest general-purpose ML models to explore and visualize datasets. Figure~\ref{fig:chemiscope-explore} shows the MD22\cite{MD22} dataset, visualized in \texttt{chemiscope} using the general-purpose featurizer based on PET-MAD with dimensionality reduction as discussed in reference~\citenum{mazitov2025massive}.

\section{Code quality and usability}

Because scientific software is not an end in itself, but rather a tool we use to study the physical world, the ease of use, ease of installation, and long term maintainability are at least as important as the set of features exposed by a given piece of software. For these reasons, \metatensor and all related software take a holistic development approach, considering both new features and performance improvement at the same time as ensuring that the resulting software is easy to use and install by anyone, regardless of which computing devices they have access to.

\metatensor and the wider ecosystem around it are developed publicly on GitHub at \url{https://github.com/metatensor}, allowing everyone to interact with the code and shape its future. Contributions of all kinds are welcome, from documentation to design work. Code contributions are automatically checked with continuous integration, making sure none of the existing tests are broken by the new contributions.

The ecosystem is aligned with the FAIR principles\cite{communityTuringWayHandbook2019,wilkinsonFAIRGuidingPrinciples2016}, with the output files also conforming to open standards: \resub{\metatensor uses NumPy's \texttt{npz} format for storage (with the \texttt{.mts} extension)}, allowing easy interoperability across languages.

Finally, a large amount of work goes into making the code available to end users in the most convenient way possible. For libraries, this means distributing it in package indexes such as PyPI\cite{pypi} for Python, conda-forge\cite{conda-forge} for Python, C and \cxx, or crates.io for Rust. Where relevant, the libraries are pre-compiled to maximize compatibility with a variety of user machines. We also provide pre-compiled packages for applications, such as \metatrain (available on PyPI and conda-forge); the metatomic-enabled version of LAMMPS (available on conda-forge, including support for multiple GPU architectures) and the metatomic-enabled versions of PLUMED (available on conda-forge).

We also support direct compilation of the code through a set of \texttt{spack}\cite{Gamblin2015} recipes for use in HPC centers, and each project uses standard build tools (\texttt{CMake} for C++, \texttt{cargo} for Rust, \texttt{setuptools} for Python) that will be familiar to advanced users wanting to manually compile the software or install development versions.

\section{Outlook}

The landscape of software for atomistic modeling has long been dominated by high-performance-oriented codes written in Fortran, C, and \cxx, primarily implementing electronic-structure methods and molecular-simulation algorithms. Many of these codes have legacies stretching back decades. Recent advances in machine learning (ML) have reshaped this equilibrium, introducing a new wave of software --- often written in more interactive languages such as Python and Julia --- characterized by fundamentally different needs for data handling, model sharing, and software interoperability. These two ecosystems, the established HPC-oriented codes and the emerging ML-driven frameworks, must increasingly work together to address modern scientific challenges. \metatensor and \metatomic address this need for interoperability in atomistic ML by introducing a new labeled-data storage and exchange format, designed specifically for ML workflows involving atomic-scale systems. \metatensor data containers are metadata-rich, sparse by design, and gradient-friendly, making them a natural interchange format between libraries that may differ greatly in maturity, architecture, and purpose. Multi-platform and multi-language compatibility is central to the design: the C API ensures seamless integration with both legacy and modern codes, regardless of programming paradigm.

The rise of ML also changes the paradigm of code sharing. Whereas traditional physical models could be reproduced from code alone, ML models additionally require sharing learned parameters. \metatomic addresses this by defining a unified container format for arbitrary atomistic ML models built using TorchScript. This allows heterogeneous models to be handled uniformly by both users and third-party libraries. An atomistic simulation engine can get access to with a wide variety of ML architectures by supporting the \metatomic interface.

Beyond core functionality, we place strong emphasis on accessibility: all parts are accompanied by clear, beginner-friendly documentation and recipes in the Atomistic Cookbook (\url{https://atomistic-cookbook.org/}), which illustrates how to combine different parts of the ecosystem in real-world workflows through practical examples. \metatensor and \metatomic are already central components in a rapidly expanding software ecosystem, enabling unique capabilities in many libraries and tools. These integrations work symbiotically: the metadata-rich, multi-platform nature of \metatensor enhances the functionality of downstream codes, which in turn broaden the scope of the overall ecosystem.

\resub{Looking ahead, we plan to integrate \metatensor with more ecosystems, including JAX, Julia, and Fortran; and to extend support to define custom models, giving users access to the tools they are used to from large ML libraries. For \metatomic, we aim to decouple it from TorchScript and add support for multiple kinds of models, still including TorchScript while also allowing to run pure Python scripts, Julia models or even native \cxx code in a shared library to be used as a model together with any of the simulation engines compatible with \metatomic.}

We envision \metatensor and its ecosystem as catalysts for a new level of usability, interoperability, and integration within atomistic modeling, empowering both users and developers to tackle the most pressing challenges in computational chemistry and materials science.

\section{Author contributions}

\newcommand{\GF}{\textbf{G.~Fraux}\xspace}
\newcommand{\FB}{\textbf{F.~Bigi}\xspace}
\newcommand{\PL}{\textbf{P.~Loche}\xspace}
\newcommand{\JAB}{\textbf{J.~W.~Abbott}\xspace}
\newcommand{\AG}{\textbf{A.~Goscinski}\xspace}
\newcommand{\DT}{\textbf{D.~Tisi}\xspace}
\newcommand{\AM}{\textbf{A.~Mazitov}\xspace}
\newcommand{\PP}{\textbf{P.~Pegolo}\xspace}
\newcommand{\SaC}{\textbf{S.~Chong}\xspace}
\newcommand{\SoC}{\textbf{S.~Chorna}\xspace}
\newcommand{\RG}{\textbf{R.~Goswami}\xspace}
\newcommand{\MC}{\textbf{M.~Ceriotti}\xspace}
\newcommand{\ML}{\textbf{M.~Langer}\xspace}
\newcommand{\MK}{\textbf{M.~Kellner}\xspace}
\newcommand{\PF}{\textbf{P.~Febrer}\xspace}

\GF led the development of \metatensor and \metatomic, with many contributions by \FB, \JAB, \PL, \AG, \DT, \PP, and \RG.
\PL and \FB led the development of \metatrain, with contributions from \GF, \AM, \SaC, \JAB, \PF, \DT, and \PP.

\FB and \GF implemented the interface between metatomic and LAMMPS and ASE. \GF implemented the interface between metatomic and PLUMED. \RG implemented the interface between metatomic and eOn. \SoC implemented the interface between metatomic and chemiscope. \FB and \MC implemented the interface between metatomic and i-PI.

\GF led the development for \texttt{featomic}, with contributions from \PL and \JAB. \ML and \FB implemented the \texttt{torch-spex} package. \PL and others wrote \texttt{torch-pme}. \FB, \MC, and others implemented \texttt{sphericart}\cite{bigi+23jcp}. \GF implemented \texttt{vesin}.

\MK created the ShiftML 3.0 model\cite{kellnerDeepLearningModel2025a} using \metatrain. \AM and \FB created the PET-MAD model\cite{mazitov2025petmad} using \metatrain. \FB created the FlashMD model\cite{bigi2025flashmd}.

All authors contributed to the writing of this article.

\begin{acknowledgments}
The authors would like to acknowledge the many supporting contributions to \metatensor, \metatomic and their ecosystem, both from members of the Laboratory of Computational Science and Modeling and the broader community.
These implementation efforts were generously supported by the NCCR MARVEL, a National Centre of Competence in Research, funded by the Swiss National Science Foundation (grant number 182892); by the Swiss Platform for Advanced Scientific Computing (PASC); by the European Research Council (ERC) under the European Union’s Horizon 2020 research and innovation programme (grant agreement No 101001890-FIAMMA); by the Swiss National Science Foundation (Project No. 200020-214879); by the European Center of Excellence MaX, Materials at the Exascale - GA No. 676598.

\end{acknowledgments}

\subsection*{Data availability}

All the software discussed in this work is available freely under an open source license. The documentation of \metatensor and \metatomic is available at \url{https://docs.metatensor.org/}. Scripts used to generate \resub{some of the figures} are available at \url{https://github.com/metatensor/ecosystem-article}.

\end{document}